\documentclass[11pt]{amsart}
\usepackage{graphics}
\usepackage{hyperref}
\font\a=cmr10
\font\p=cmr9
\font\gr=cmr10 scaled 1200
\def \dd{{\rm d}}
\def \DD{{\rm D}}
\numberwithin{equation}{section}
\begin{document}
\title[Reinstating Schwarzschild's original manifold]
{Reinstating Schwarzschild's original manifold and
its singularity}%
\author{Salvatore Antoci}%
\address{Dipartimento di Fisica ``A. Volta'' and C.N.R., Pavia, Italia}%
\email{Antoci@matsci.unipv.it}%
\author{Dierck-Ekkehard  Liebscher}%
\address{Astrophysikalisches Institut Potsdam, Potsdam, Deutschland}%
\email{deliebscher@aip.de}%
\maketitle

{\a Abstract. A review of results about this paradigmatic
solution\footnote{The notion solution \textit{includes} the topology of
space-time in our context. Different solutions may be \textit{locally} isometric.}
to the field equations of Einstein's theory of general relativity
is proposed. Firstly, an introductory note of historical character
explains the difference between the original Schwarzschild's
solution and the ``Schwarzschild solution'' of all the books and
the research papers, that is due essentially
to Hilbert, as well as the origin of the misnomer.\par
The viability of Hilbert's solution
as a model for the spherically symmetric field of a ``Massenpunkt"
is then scrutinised. It is proved that Hilbert's solution
contains two main defects. In a fundamental paper written in 1950,
J.L. Synge set two postulates that the geodesic paths of a given
metric must satisfy in order to comply with our basic ideas on
time, namely the postulate of \textit{ order} and the
\textit{non-circuital} postulate. It is shown that neither Hilbert's
solution, nor the equivalent metrics that can be obtained
from the latter with a coordinate tranformation that is regular
and one-to-one everywhere except on the Schwarzschild surface
can obey both Synge's postulates. Therefore they do not possess a
consistent arrow of time, and the only way for obviating this defect
is through a change of topology. The true raison d'\^etre of the
Kruskal maximal extension with its odd doubling and bifurcate horizon
stays just in its capability to produce the needed change of topology,
that can be demonstrated through a constructive
cut-and-past procedure applied to two Hilbert space-times.\par
The second main defect of Hilbert's space-time is constituted by the
existence of an invariant, local, intrinsic quantity
with a simple operational interpretation that diverges
when it is calculated at a position closer and closer to
Schwarzschild's surface, i.e. at an internal position
in Hilbert's metric. The diverging quantity is the norm of the
four-acceleration of a test particle whose worldline is
the unique orbit of absolute rest defined, through a given
event, by the unique timelike, hypersurface orthogonal Killing vector.
It is an intrinsic quantity, whose local definition only requires
the knowledge of the metric and of its derivatives at a given
event, just like it happens with the polynomial invariants built with the
Riemann tensor and with its covariant derivatives. The regularity
of the latter invariants at a given event has been considered by many a relativist
like a ``rule of thumb'' proof of regularity for the manifold at that
event, in the persistent lack of a satisfactory definition of local singularity
in general relativity.\par
    The divergence of the above mentioned norm of the
four-acceleration, i.e. of the first curvature of the
worldline, is a geometric fact. It can be proved however with an
exact argument, relying on a two-body solution found by Bach,
that a physical quantity, the norm of the force per unit mass
exerted on a test particle in order to keep it
on the orbit of absolute rest,
is equal to the norm of the four-acceleration, hence it diverges
too on approaching Schwarzschild's surface.\par
     We claim that the r\^ole and interpretation of topological
differences between partly isometrical manifolds as well as that
of the singularities is not really settled, in particular that the
Schwarzschild solution and its topological relatives are in more
ways singular than the invariants of the Riemann tensor
indicate.\par
Due to these facts we assert that the topology chosen by
Schwarzschild should be taken as a serious alternative
to the commonly used Hilbert or Kruskal topologies.}

\section{Introduction: Schwarzschild's original solution and
the ``Schwarzschild solution''}
The content of this review would be hardly understandable without
a proviso of historical character: Schwarzschild's original
solution, as undisputably testified by Schwarzschild's ``Massenpunkt''
paper \cite{Schwarzschild1916}, describes a manifold that is different from the one
defined by the solution that goes under the name of
Schwarzschild in practically all the books and the research
articles written by the relativists in nearly nine decades. That
solution must be instead credited to Hilbert \cite{Abrams1989}.
The readers should not be
induced by this assertion into believing that it is our intention to
deprive Schwarzschild of the merit of his discovery, and to attribute
it to the later work by Hilbert. It is not so: an accurate reading of
Schwarzschild's paper and of the momentous Communication by Hilbert
\cite{Hilbert1917} shows in fact that, while Schwarzschild's
derivation of the original solution is mathematically flawless,
Hilbert's rederivation contains an error. Due to this overlooked flaw,
Hilbert's manifold
happened to include Schwarzschild's manifold, but by chance it
resulted not to be in one-to-one correspondence with it. This fact was deemed
rather irrelevant by Hilbert, but soon developed into a conundrum that
puzzled theoreticians like Marcel Brillouin \cite{Brillouin1923}, and had to become
of crucial importance more than forty years later. In fact the birth of
the black hole idea, as noted for the first time by Abrams
\cite{Abrams1989}, can be considered just a legacy of Hilbert's magnanimity.\par
Schwarzschild's paper, in which it is reported the first derivation
of the field of a ``Massenpunkt'' according to general relativity,
is an impressive example of mathematical precision and clearness
of exposition. For the sake of definiteness in the subsequent discussion,
an English translation \cite{Schwarzschild2003} of that paper
is provided in Appendix \ref{A}.\par
On reading Schwarzschild's paper, one immediately notes \cite{Note2003}
that the equations he set out to solve in the special case of a
static, spherically symmetric field are not the final equations
of general relativity \cite{Einstein1915c, Hilbert1915} found by Einstein
and by Hilbert, specialised for the case of vacuum. In fact,
equations (\ref{A4}) and (\ref{A5})
are the vacuum equations for the next-to-last version of the
theory \cite{Einstein1915a} that Einstein communicated
to the Prussian Academy of Sciences on November 11th, 1915.
These equations yield, in vacuo, the same solutions as the equations
of the final theory of November 25th, but their covariance is limited
to the unimodular transformations.\par
    Due to this fact, it was highly inconvenient for Schwarzschild
to avail, as symmetry adapted coordinates, of the usual spherical
polar coordinates $r$, $\vartheta$, $\varphi$, $t$, since the
functional determinant of the transformation from Cartesian to
such polar coordinates is $r^2\sin\vartheta\ne 1$; he had rather
adopting \textit{ polar coordinates with the determinant 1}, defined
by equation (\ref{A7}). Using these coordinates $x_i$,
the square of the symmetry adapted line element $\dd s^2$ is written
by Schwarzschild, in equation (\ref{A9}),
by means of $f_1$, $f_2=f_3$, $f_4$, i.e. three independent functions
of the radial coordinate $x_1$. These functions must fulfill the
conditions enumerated just after that equation: Minkowskian behaviour at the
spatial infinity, field equations, inclusive of the equation
of the determinant, and continuity everywhere, except at the
origin.\par
When these conditions are obeyed, with the exception of the
last one, the three independent functions $f_1$, $f_2$, $f_4$
were found by Schwarzschild to be given by equations
(\ref{A12}), (\ref{A10}) and (\ref{A11}) respectively. They
contain two independent integration constants, $\alpha$
and $\rho$. By appealing to Newton, Schwarzschild
obtained that $\alpha$ had, in his units, just twice the value of the active
gravitational mass. The remaining constant $\rho$ was instead
determined by Schwarzschild by imposing his last condition,
namely the continuity of the components of the metric.
The function $f_1$ is in fact discontinuous when
$3x_1=\alpha^3-\rho$. For a given $\alpha$, the choice of
a given value for $\rho$ changes the position of the
discontinuity in the interval of  definition $[0,+\infty[$ of the
radial coordinate $x_1$. It can also bring the discontinuity
outside it. Therefore the choice of $\rho$ fixes the very choice of
the manifold that describes the field of a ``Massenpunkt''.
The requirement of continuity everywhere, except at the origin,
led Schwarzschild to set $\rho=\alpha^3$, i.e. to locate
at the inner border of the manifold the discontinuity that would
have been named, after him, ``the singularity at the Schwarzschild
radius''.\par
Let us compare now Schwarzschild's determination of the manifold
representing the static, spherically symmetric gravitational field
with the later derivation done by Hilbert \cite{Hilbert1917}. For
ease of comparison, an English translation for the relevant part
of the second of Hilbert's Communications entitled ``Die Grundlagen der
Physik'' is reproduced in Appendix~\ref{B}. Hilbert could avail of the
generally covariant, final version of the theory, that he himself had
contributed to establish with his first Communication \cite{Hilbert1915}
bearing the same title. Hence he could adopt the
usual spherical polar coordinates and write the line element
for a spherically symmetric, static $g_{\mu\nu}$ field like in
expression (\ref{B42}). Thereby the line element is made to
depend on three arbitrary functions
$F(r)$, $G(r)$, $H(r)$, where $r$ is a radial coordinate, whose
range, in keeping with its definition in term of the Cartesian
coordinates $w_i$, is given by $0\le r< +\infty$. But, while
Schwarzschild's choice of a line element depending on three
arbitrary functions was not redundant, since he had to fulfill
both the field equations and the condition of the determinant,
Hilbert's line element, after satisfying the field equations,
would still contain one arbitrary function. Therefore Hilbert
chose to fix one of the three functions $F(r)$, $G(r)$, $H(r)$ by defining a
new radial coordinate $r^{\ast}$ such that $r^{\ast}=\sqrt{G(r)}$.
Then, of course, he was entitled to drop the asterisk and
write the line element with two unknown functions  $M(r)$ and
$W(r)$, like in expression (\ref{B43}), that
has become the canonical starting point for all the textbook
derivations of the ``Schwarzschild solution''. As correctly
remarked by Abrams \cite{Abrams1989}, what neither Hilbert
nor,  in his footsteps, the subsequent generations of relativists
were entitled to do was assuming without justification that the physical range of the
new radial coordinate is still $0\le r< +\infty$. In fact,
this is equivalent to inadvertently make the restrictive choice
$G(0)=0$ in expression (\ref{B42}).\par
The new radial coordinate allowed Hilbert
to avail of a straightforward method of solution, based on the bright,
although mathematically unwarranted shortcut: writing the symmetry
restricted action for the problem under study and applying
a symmetry restricted variational principle to it; the correctess
of the resulting solution is thereby not ensured, and must be checked
afterwards. In this way Hilbert eventually found the line element
of equation (\ref{B45}), that,
at variance with Schwarzschild's result, depends on just one
integration constant, $\alpha$, again interpreted as twice the
active gravitational mass. The discontinuity in the coefficient
of $\dd r^2$, that is the counterpart of the discontinuity
exhibited by the function $f_1(x_1)$ in Schwarzschild's equation
(\ref{A12}), in Hilbert's solution is fixed at $r=\alpha$, because
in Hilbert's calculation there is no free integration constant,
like the $\rho$ of Schwarzschild, to move it around. This is the
consequence of the unduly restrictive choice $G(0)=0$, i.e. of Hilbert's
lapse, and by no means a necessary consequence of Einstein's
equations. Einstein's equations do not fix the topology, and additional
reasoning is necessary to do so.\par
Nevertheless, Schwarzschild's original solution and manifold,
that were the result of a mathematically correct procedure and of
the pondered choice of the integration constant $\rho$, went
soon forgotten. Hilbert's solution and manifold, that were on the
contrary the consequence of the inadvertent fixing of an integration constant,
were instead handed down to the posterity as the unique
``Schwarzschild solution''. A contribution to the rooting of the
misnomer undoubtedly came from Schwarzschild's premature death.
But the main responsibility for it should be attributed to Hilbert
himself, given his exceptional, well deserved standing
in the community of the mathematicians, astronomers and
physicists of his time, and of ours.\par
In fact, while dismissing, in a short footnote\footnote{see Hilbert's
footnote at the end of Appendix \ref{B}.}, Schwarzschild's
procedure of removing the discontinuity to the origin as
``not advisable'', Hilbert attributed to
Schwarzschild the finding of his own solution (\ref{B45}),
hence of the manifold inadvertently chosen by himself, the
one for which $0\le r< +\infty$, with its two singularities
located at $r=0$ and at $r=\alpha$ in what the relativists would
have called ``Schwarzschild coordinates''.\par
However such features, that would have so much bothered generations of
scholars, were only of very marginal interest for the great Hilbert.
As it appears from the enthusiastic, last words of his Communication \cite{Hilbert1915}
of 1915, he harbored the firm conviction that, by starting from
just two very simple axioms, thanks to the powerful instruments
constituted by the calculus of variations and by the theory of
invariants, he had succeded in embodying both Einstein's new conceptions
about gravitation and Mie's new ideas about electrodynamics
\cite{Mie1912, Mie1913} into a mathematical structure of
lasting value. In Hilbert expectations, only the full theory
would have been capable to provide an immediate representation of
reality, at any scale, through everywhere regular solutions. Hence, one
should not worry if the first, very partial
soundings into the exact mathematical content of the theory,
done by neglecting the fundamental ingredient of the electromagnetic
field, exhibited, together with the confirmation of Einstein's
achievement \cite{Einstein15b} about the perihelion of Mercury, a singular
behaviour of difficult interpretation.
\section{The wrong arrow of time of Hilbert's manifold is at
the origin of the Kruskal extension}
It is not here the place to recall in detail how it happened that the relativists
gradually abandoned the attitude to look in general relativity for the
simile of the notion of singularity that had been so useful
in the mathematical physics of the past. That ideal conception is
well reflected in the very notion of regularity of the interval
and of the metric that Hilbert provided just at the end of his
derivation of the ``Schwarzschild solution''. According to Hilbert's
definition, the singularities exhibited by the components of the
metric both at $r=0$ and at $r=\alpha=2m$ were true singularities of
the field $g_{\mu\nu}$, because they
could not be erased by invertible, one-to-one transformations
of coordinates. The notions of regularity and of singularity set by
Hilbert went unchallenged, with some notable exceptions, till
the half of the past century, when they found also a sharpened
definition \cite{Lichnerowicz1955} in the book by Lichnerowicz.
It should be remarked here that Hilbert's definitions of regularity
and of allowable coordinate transformations not only represented
the extension to the new theory of
time-honoured conceptions of mathematical physics, but were in
keeping with Einstein's own ideas on the meaning of general covariance
\cite{Einstein1916}. According to Einstein, it was in fact necessary to divest
space and time coordinates of the last residuum of physical objectivity,
but for one aspect. For him, all the assertions of physics could be reduced
in the last instance to assertions about events, physically embodied
by spacetime coincidences between particles. The introduction of spacetime
coordinates is a convenient instrument for reckoning such
coincidences. But for the coordinates to absolve
their residual physical function, it is necessary
that the transformations of coordinates
do preserve the individuality of the single event. There is
therefore one sound physical reason why the allowed transformations
must be invertible and one-to-one, as required by Hilbert.\par
Einstein's conception of the ``Bezugmolluske'' and the connected
definition of regularity by Hilbert would have resisted the
lapse of the decades, if the manifolds of general relativity had
been truly Riemannian, rather than pseudo-Riemannian, like they
happen to be, and if the singular surface at $r=\alpha$ in the
Hilbert solution had not had null lines as generators. This is why
the doubt first arose about the true nature of the ``singularity
at the Schwarzschild radius''. Along each one of the generators
it is $\dd s^2=0$. Does it mean that we have to do with a light path, or
does it mean that this generator is not a worldline, but just
one and the same event, misrepresented in a
coordinate system in which an inadequate choice of the coordinate
chart has created, together with the singularity of the metric,
the illusion of the presence of a light line? If so, why not
try to find, through some disallowed coordinate transformation,
appropriately singular at $r=\alpha$, a new coordinate chart in which
the light line, for all the finite values of the Hilbert coordinate time $t$,
would become associated with just one value of the
new coordinate time, and the metric would
become regular at $r=\alpha$?\par
After preliminary attempts that
succeeded in the second task, but not in the first one
\cite{Eddington1924, Finkelstein1958, Lemaitre1933}, both tasks
were eventually accomplished together by Synge \cite{Synge1950}
in a ground-breaking, now nearly forgotten paper. In its footsteps
came the results by Fronsdal, Kruskal and Szekeres
\cite{Fronsdal1959, Kruskal1960, Szekeres1960} about the so-called maximal
extension of the Schwarzschild manifold. Of course, in
order to accomplish both tasks, Synge had to infringe the rules
about the admissible transformations set by Einstein and Hilbert. Synge's ideas
about the definition of singularities in general relativity are worth
mentioning, because they were very clear, and pessimistic. With some reason,
one may add, because the situation, as we shall see in the sequel, has not
greatly improved since the time Synge's paper \cite{Synge1950} appeared
in print. Synge wrote:
\begin{quotation}
Obviously, before we talk of singularities at all, we should
define them. Adequate definitions should be invariant, but there
are difficulties here which may not appear on the surface. It is
true that some of these difficulties are overcome by a limitation
to regular transformations, but it is precisely the non-regular
transformations which are interesting. Thus we must satisfy
ourselves for the present with definitions dependent on the
coordinate system employed.
\end{quotation}
And then he provided his coordinate-dependent definition:
\begin{quotation}
\textit{ Definition.} A form $g_{mn}\dd x^m\dd x^n$ has a
\textit{component singularity} at a point $x^r=a^r$ which lies in a region
$R$ of the representative space $V_4$ or on the boundary of
$R$, if one or more of the components $g_{mn}$ has no unique
finite limit, independent of path, as we approach $a^r$ through
$R$; and it has a \textit{ determinantal singularity} if the
determinant $g$ of $g_{mn}$ has not a unique finite \textit{ non-zero}
limit as we approach $a^r$ through $R$.
\end{quotation}

In keeping with this definition, one may assert that Synge and his followers had
succeeded, by an appropriately singular transformation,
in removing the singularity that the metric components exhibit
at $r=\alpha$. They had also succeeded, by the same singular
transformation, in reducing the points with
finite $t$ coordinate on each one of the generators of the
singular surface $r=\alpha=2m$ to have one and the same coordinate time
in the new chart. As a consequence of this
second achievement, it became possible to draw and explore
mathematically how the timelike and the null
geodesics prevailing in the regions of Hilbert's manifold with $r<\alpha$ and with
$r>\alpha$ respectively happen to connect smoothly at $r=\alpha$.
Such a connection could not be properly explored
in the original coordinates adopted by Hilbert, because
in these coordinates the connection of the geodesics occurs when
$t\rightarrow\pm\infty$, as it is hinted by Figure (1a).
\begin{figure}[ht]
\includegraphics{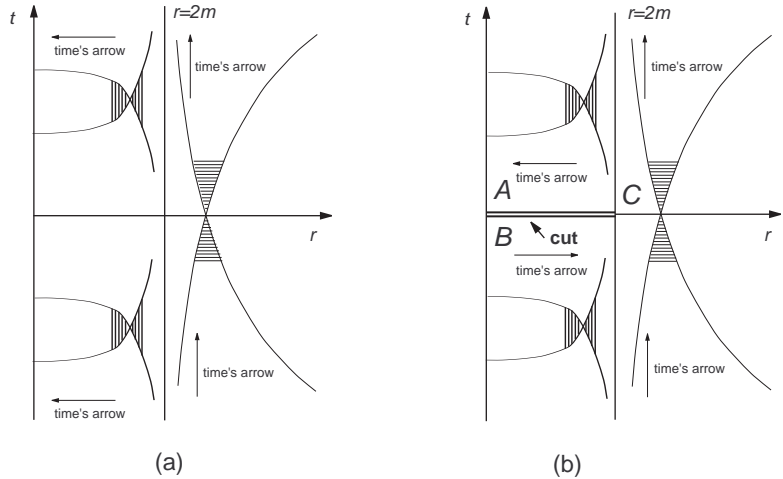}
\caption{\p (a): Representation of Hilbert's manifold in the $r$, $t$ plane.
Light cones are drawn both in the inner region and in the outer
region. Time arrows are schematically drawn in keeping with the
non-circuital postulate. The postulate of order is thereby
violated. (b) A cut is made in Hilbert's manifold. The
topologically different manifold obtained in this way
allows for a consistent drawing of the time arrow.}
\end{figure}
\par
One can now notice that Synge and his followers, with their singular
coordinate transformations, had reached a third achievement too.
This further result is usually given scarce or no relevance at all in the literature,
although e.g. Rindler did not forget to make a fugitive mention of it
in his book \cite{Rindler2001}. In order to appreciate its full value,
one has rather resorting once more to Synge, were a detailed discussion
of the issue of the time arrow can be found \cite{Synge1950}.
According to Synge, a manifold meant
to be a model of physical reality must fulfill two postulates.
One of them is the \textit{ postulate of order}, according to which the parameter of
proper time along a timelike geodesic must always either decrease
or increase; the sense along which it is assumed to increase defines the sense
of the travel from past to future, namely the time arrow. Since the geodesic
equation is quadratic in the line element, fixing the time arrow
of the individual geodesic is a matter of choice. The
second postulate deals with our ideas of causation, and
establishes a relation between the time arrows of neighbouring
geodesics. Synge calls it the \textit{ non circuital postulate}.
It asserts that \textit{ there cannot exist in space-time a closed
loop of time-like geodesics around which we may travel
always following the sense of the time-arrow}.\par
Synge shows in detail \cite{Synge1950} that the time arrow can be drawn in keeping
with the aforementioned postulates in the manifold that he
obtained from the Hilbert manifold with his singular coordinate
transformation; the same property obviously holds in the
Kruskal-Szekeres manifold too. Does it hold also in the Hilbert
manifold? A glance to Figure (1a) is sufficient to gather that
this is not the case. The arrow of time can be drawn correctly
in the submanifold with $r>\alpha$, i.e. in Schwarzschild's
original manifold, and separately in the inner submanifold with
$r<\alpha$. A consistent drawing of the arrow of time, in keeping
with both postulates, is however impossible in Hilbert's manifold
as a whole. This is an intrinsic flaw of the latter
manifold; it has nothing to do either with the fact that in Hilbert's
chart the metric is not defined at $r=\alpha$, or with the
fact that the timelike geodesics appear to cross the Schwarzschild surface
at the coordinate time $t=\pm\infty$; it is a flaw that cannot be remedied
by \textit{ any} coordinate transformation, however singular at $r=\alpha$, but
one-to-one elsewhere.\par
The only known ways to overcome this flaw are either by getting rid of
the inner region, thereby reinstating the original manifold \cite{Schwarzschild1916},
deliberately chosen by Schwarz\-schild as a model for the gravitational field
of a material particle, or by completely renouncing the one-to-one
injunction on the coordinate transformations set, on physical
grounds, by Einstein and by Hilbert.\par
The second alternative is the one chosen by Synge and his
followers: in fact, not only Schwarzschild's original manifold, but
also Kruskal's manifold avoids the flaw of the arrow of time
present in Hilbert's manifold. Moreover, it appears to preserve its inner
region. However, it does so by a coordinate transformation that duplicates the
original manifold and alters its topology, in a way that
is best explained, rather than by looking at the equations for the
transformation, by a straighforward cut-and-paste procedure
applied to two Hilbert manifolds.\par
This cut-and-past procedure can be realised in infinite
ways, all entailing an alteration of topology. One of them is
accounted for in the sequence drawn in Figures (1b), (2a) and (2b).
\begin{figure}[ht]
\includegraphics{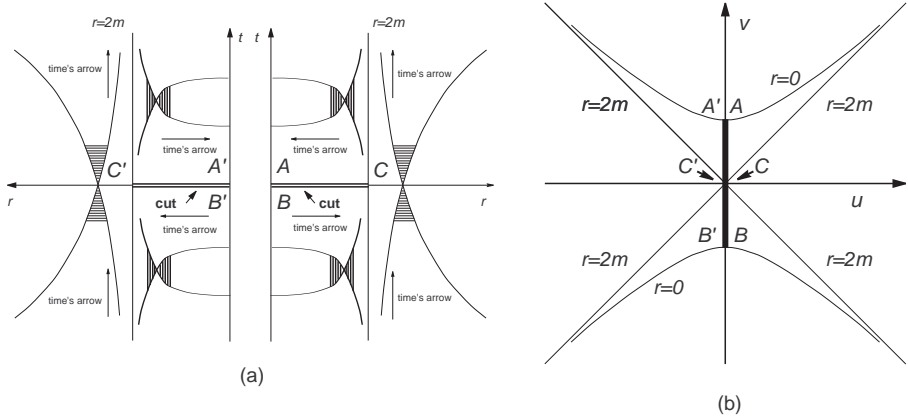}
\caption{\p (a) two manifolds, equal to the cut Hilbert manifold
of Figure (1b) are juxtaposed for suturing; (b) sewing together
the edges $ACB$ and $A'C'B'$ yields Kruskal's manifold.}
\end{figure}
In Figure (1b) the inner region of the Hilbert manifold of Figure
(1a) is cut along the line $AC$. The resulting manifold is
topologically inequivalent to the Hilbert manifold. The
topological alteration already allows to draw
the arrows of time in keeping with Synge's
postulates but, due to the existence of the border
$ACB$, the new manifold is evidently unphysical. However, if one
takes two manifolds identical to the one of Figure (1b),
juxtaposes them as it is shown in Figure (2a),
and eventually sews together the borders $ACB$ and $A'C'B'$
like in Figure (2b), one obtains Kruskal's manifold, and
ascertains that the arrows of time inherited
from the two component manifolds with the cut still obey both
Synge's postulates.\par
Therefore, rather than discarding it, like one does with the Schwarzschild
manifold, one can save the inner region of Hilbert's manifold
at the cost of changing the topology of the latter with a cut,
that is then remedied by the doubling intrinsic to the
Kruskal-Szekeres manifold. The mathematical beauty of the latter is
beyond question. Its physical usefulness much less so,
if, after having examined at length the properties of the Kruskal
metric, textbooks usually end the discussion by asserting
that this indissoluble union of a white and a black hole
hardly has anything to do with some entity really present in Nature.
It is usually added that black holes should be meant as the final result of
the process of collapse, in which only part of the Kruskal manifold
would be involved.
It is however evident that entering such arguments means
abandoning the discussion of actually existing, well scrutinised vacuum solutions
to the field equations of general relativity, and venturing,
despite the huge amount of work done on the subject, in uncharted
waters.\par
Anyway, when confronted with the Kruskal manifold, with its odd doubling and
its bifurcate horizon, we eventually gather how serious is the flaw
of the arrow of time in Hilbert's manifold,
if one could conceive resorting to such extreme surgical means in
order to remove just that flaw.
\section{An invariant, local, intrinsic quantity that diverges at the
Schwarzschild surface}
We have called Synge's ideas about the issue of the definition of
a singularity in general relativity clear and pessimistic. His
coordinate-dependent definition is in fact a clear mirror of the
difficulties that he met with. Of course he would have had
rather availing of an invariant definition of singularity, because
in such a way the problem of defining what coordinate
transformations were allowable would have become much less challenging.
In fact, a scalar would maintain its value, no matter whether
the transformations of coordinates should be limited to the ones
allowed for by Einstein and Hilbert, or whether Synge's
``interesting'' transformations, namely nonregular ones, thereby
capable of canceling the singular behaviour of the metric at $r=\alpha$ in
Hilbert's solution, could be allowed too. However,
no general enough, physically satisfactory, invariant definition
of singularity was available to him. A posteriori one must recognise that
Synge's pessimism was justified, because for decades the problem has been
ingeniously tackled by several authors
(see for instance \cite{Geroch1968a, Geroch1968b, Schmidt1971, GKP1972}
\cite{ES1977, Thorpe1977, GLW1982, SS1994}) who proposed several
solutions but, by the admission \cite{GLW1982} of some of them, it is also
possible that ``there may not exist any useful, generally applicable
notion of the singular boundary of a space-time''.\par
Synge's coordinate-dependent definition, on the other side, cannot be a
satisfactory answer if the nonregular transformations are decreed
legitimate, for then the locus $r=\alpha$ is either singular or
regular according to whether it is looked at in the Hilbert chart
rather than in the Eddington-Finkelstein
\cite{Eddington1924, Finkelstein1958} or Kruskal charts respectively.\par
In the persistent lack of a physically satisfactory, general, invariant
definition of singularity, and due also to the growing awareness
that Synge's \cite{Synge1966} identification
\textit{ ``gravitational field = curvature of space-time''}
had to be substituted for Einstein's old idea\footnote{see
\cite{Einstein1916}, p. 802:
``Verschwinden die $\Gamma^\tau_{\mu\nu}$, so bewegt sich der Punkt
geradlinig und gleichf\"ormig; diese Gr\"o\ss en bedingen also die
Abweichung der Bewegung von der Gleich\-f\"ormigkeit. Sie sind die
Komponenten des Gravitationsfeldes.''} that in general relativity
the gravitational field was represented by the Christoffel
symbols, the scalar invariants
built with the Riemann tensor\footnote{more precisely, the scalar
polynomial expressions built with the metric $g_{ab}$, with the Levi-Civita
symbol $\epsilon^{abcd}$, with the Riemann tensor $R_{abcd}$
and with its covariant derivatives.}, whose singularity is
certainly sufficient for proving the singularity of a manifold,
began to be used, for lack of a better alternative, as invariant, local,
intrinsic indicators of its regularity too.\par
In the case of the Kruskal manifold, the latter scalar invariants
provided the same verdicts about singularity and \textit{ ad ignorantiam}
regularity as Synge's coordinate dependent definitions applied in
the Kruskal chart: the singularity at $r=\alpha$ is fictitious,
due to the choice of the coordinates, while the singularity at $r=0$
is a true one, intrinsic to the manifold.\par
There is however one argument that suggests that such invariants
might not be as trustful indicators of regularity as they are
generally believed to be. Metrics endowed with spherical symmetry
are indeed very special items. When availing of a given metric, endowed
with a high degree of symmetry, as a model of physical reality, one would
like not to incur the following, perplexing situation: as soon as
a certain symmetry of the metric is lifted, one of
its essential properties, as measured by some
scalar quantity, changes abruptly,
no matter how small it is the deformation that destroys the original symmetry.\par
Let us consider what is generally called a static spacetime, namely a
pseudoRiemannian manifold that admits a timelike Killing vector that is hypersurface
orthogonal; this Killing vector $\xi_i$ fulfills the equations:
\begin{equation}\label{1}
\xi_i\xi^i>0, \ \ \xi_{i;k}+\xi_{k;i}=0, \ \ \xi_{[i}\xi_{k,l]}=0.
\end{equation}
In symmetry adapted coordinates  $x^0$, $x^1$, $x^2$, $x^3$,
the line element of a static spacetime reads
\begin{equation}\label{2}
\dd s^2=\sum_{\alpha,\beta=1}^3
g_{\alpha\beta}(x^1,x^2,x^3)\dd x^{\alpha}\dd x^{\beta}
+V^2(\dd x^0)^2, \ \ V=V(x^1,x^2,x^3).
\end{equation}
Let us assume the line element to enjoy the following
properties:
\begin{itemize}
\item the metric is a vacuum solution of Einstein's field
equations, regular and Min\-kowskian at spatial infinity;
\item the equipotential surfaces $V=C=$ const.$>0$, $x^0=$ const.
are regular, simply connected, closed 2-surfaces;
\item the intrinsic geometry of the 2-surfaces
$V=C, x^0=$ const., tends to that of a closed regular 2-surface of finite
area in the limit $C\rightarrow 0$;
\item the Kretschmann scalar $R_{iklm}R^{iklm}$ is everywhere
finite.
\end{itemize}
Then a surprising theorem, found \cite{Israel1967a} by Israel,
asserts that the only spacetime that fulfills the hypotheses
enumerated above is Schwarzschild's original manifold
\cite{Schwarzschild1916}.
Of course, Israel's theorem does not say which of the
hypotheses happens to fail when the static spacetime
is not Schwarzschild's original manifold; this detail can
be learned on a case by case basis.
One interesting example is provided by the so called gamma metric.\par
The latter is one of the axially symmetric, static vacuum
solutions \cite{BW1922} calculated in 1922 by Bach, who availed of
the general method of solution \cite{Weyl1917, Levi-Civita1919}
found by Weyl and by Levi-Civita.
Despite the nonlinear structure of Einstein's equations, Weyl
succeeded in reducing the axially symmetric, static problem to
quadratures through the introduction of his ``canonical
cylindrical coordinates''. Let $x^0=t$ be the time coordinate,
while $x^1=z$, $x^2=r$ are the coordinates in a meridian half-plane,
and $x^3=\varphi$ is the azimuth of such a half-plane; the adoption
of Weyl's canonical coordinates allows writing the line
element of a static, axially symmetric field \textit{ in vacuo} as:
\begin{equation}\label{3}
\dd s^2=e^{2\psi}dt^2-\dd\sigma^2,\;e^{2\psi}\dd\sigma^2
=r^2\dd\varphi^2+e^{2\gamma}(\dd r^2+\dd z^2);
\end{equation}
the two functions $\psi$ and $\gamma$ depend only on $z$ and $r$.
Remarkably enough, in order to provide a solution to Einstein's equations,
$\psi$ must fulfill the ``Newtonian potential'' equation
\begin{equation}\label{4}
\Delta\psi=\frac{1}{r}\left\{\frac{\partial(r\psi_z)}
{\partial z}
+\frac{\partial(r\psi_r)}{\partial r}\right\}=0,
\end{equation}
where $\psi_z$, $\psi_r$ are the derivatives of $\psi$ with respect to $z$ and to
$r$ respectively, while $\gamma$ is obtained by solving the system
\begin{equation}\label{5}
\gamma_z=2r\psi_z\psi_r,\;\gamma_r=r(\psi^2_r-\psi^2_z);
\end{equation}
due to equation (\ref{4})
\begin{equation}\label{6}
\dd\gamma=2r\psi_z\psi_r\dd z+r(\psi^2_r-\psi^2_z)\dd r
\end{equation}
happens to be an exact differential.\par
The particular Bach's metric we are interested in is defined by
choosing for $\psi$ the potential that, in Weyl's ``Bildraum'',
is produced by a thin massive rod of constant linear density
$\sigma=k/2$ lying on the symmetry axis, say, between $z=z_2=-l$ and
$z=z_1=l$, that acts as source for the ``Newtonian potential''
of  equation (\ref{4}). One then finds:
\begin{equation}\label{7}
\psi=\frac{k}{2}\ln\frac{r_1+r_2-2l}{r_1+r_2+2l},
\end{equation}
where
\begin{equation}\label{8}
r_i=[r^2+(z-z_i)^2]^{1/2},\;i=1,2.
\end{equation}
By integrating equations (\ref{5}) and by adjusting an integration
constant so that $\gamma$ vanish at the spatial infinity one obtains:
\begin{equation}\label{9}
\gamma=\frac{k^2}{2}\ln\frac{(r_1+r_2)^2-4l^2}{4r_1r_2}.
\end{equation}
The resulting metric is asymptotically flat at spatial infinity
and its components are everywhere regular, with the exception of
the segment of the symmetry axis for which $z_2\leq z\leq z_1$,
for any choice of the parameters $l$ and $k$, assumed here
to be positive.\par
It may be convenient \cite{Zipoy1966} to express the line element
in spheroidal coordinates by performing, in the meridian
half-plane, the coordinate transformation \cite{CJ1973}:
\begin{equation}\label{10}
\varrho=\frac{1}{2}(r_1+r_2+2l),\;\cos\vartheta=\frac{r_2-r_1}{2l}.
\end{equation}
Then the interval takes the form
\begin{eqnarray}\label{11}
&ds^2=\left(1 - \frac{2l}{\varrho}\right)^k\dd t^2\\\nonumber
&-\left(1 - \frac{2l}{\varrho}\right)^{-k}
\left[\left(\frac{\Delta}{\Sigma}\right)^{k^2-1}\dd\varrho^2
+\frac{\Delta^{k^2}}{\Sigma^{k^2-1}}\dd\vartheta^2
+\Delta\sin^2\vartheta\dd\varphi^2\right],
\end{eqnarray}
where
\begin{equation}\label{12}
\Delta=\varrho^2 - 2l\varrho,\;
\Sigma=\varrho^2-2l\varrho+l^2\sin^2\vartheta.
\end{equation}
Therefore, provided that $k\ne 0$, the component $g_{00}=V^2$ of
this static metric exhibits closed, simply connected
2-surfaces on which $V=C$, and
$C\rightarrow 0$ when $\varrho\rightarrow 2l$, in
partial fulfillment of the hypotheses set by Israel.
We note that when $k=1$ the metric reduces to Schwarzschild's
spherically symmetric one. It does so in the strict sense:
it is in one-to-one correspondence with
Schwarzschild's original solution \cite{Schwarzschild1916},
not with Hilbert's manifold \cite{Hilbert1917}.
The latter would be retrieved from (\ref{11})
with $k=1$ if the radial coordinate $\varrho$ were allowed the range
$0\le\varrho<\infty$ while, due to (\ref{10}), the allowed
values of $\varrho$ are presently in the range $2l\le\varrho<\infty$.
Therefore, when $k=1$ the manifold fulfills all the hypotheses
set by Israel.\par
Let us now explore the behaviour, as a function of $k$, of the Kretschmann
scalar $K\equiv R_{iklm}R^{iklm}$ of the gamma metric.
The calculation and the study of this scalar has been done long
ago \cite{CJ1973} by Cooperstock and Junevicus, and was recently
repeated by Virbhadra. We quote here his result \cite{Virbhadra1996},
expressed with spheroidal coordinates:
\begin{equation}\label{13}
K=\frac{16l^2k^2N}
{\varrho^{2(k^2+k+1)}(\varrho-2l)^{2(k^2-k+1)}\Sigma^{3-2k^2} },
\end{equation}
with
\begin{eqnarray}\label{14}
N&=l^2 \sin^2\theta\left[3lk(k^2+1)(l-\varrho)
+k^2(4l^2-6l\varrho+3\varrho^2)+l^2(k^4+1)\right]\nonumber\\
&+3\varrho[(k+1)l-\varrho]^2(\varrho-2l).
\end{eqnarray}
We do not need a minute analysis of the function
$K(\rho,\vartheta,k)$ for noting that, when $\rho\rightarrow 2l$,
the Kretschmann scalar is a discontinuous parameter in the set
of the gamma metrics. It is sufficient to
examine its behaviour in the neighbourhood of $k=1$, for
arbitrarily small axially symmetric deviations from spherical
symmetry. By studying the zeroes of both the numerator and the
denominator of (\ref{13}) one is then confronted with the
following situation. For all the values of $\vartheta$, the
denominator vanishes when $\rho\rightarrow 2l$, while the values
of $\varrho$ at which the zeroes of the numerator occur depend on
both $k$ and $\vartheta$. Hence in the neighbourhood of
$k=1$ the Kretschmann scalar always diverges \cite{CJ1973} for
$\rho\rightarrow 2l$, provided that $k\neq1$. When $k=1$,
and the metric reduces to Schwarzschild's one, both the
numerator and the denominator tend to zero as $\rho\rightarrow 2l$
for all $\vartheta$, and they do so in such a way that
the limit value of the Kretschmann scalar happens to be finite at
the ``Schwarzschild radius''. Therefore, in the particular case of the gamma metric,
Israel's theorem leads to this occurrence: the slightest deviation
from the spherical symmetry of Schwarzschild's manifold destroys the
regularity of the Kretschmann scalar for $\rho\rightarrow 2l$.
This is a troublesome result,
because the latter regularity is a necessary condition for the extension of
the Schwarzschild manifold to the inner region. The above
mentioned slightest deviation from the spherical symmetry
renders the procedure of extension impossible.\par
One may conclude, like Israel did \cite{Israel1967b} in a worried letter to
``Nature'', that ``models with exact spherical symmetry
possess idiosyncrasies which render them dangerous, and perhaps
misleading, as a basis for induction''.
One may instead wonder whether, at least in this case of
static manifolds, one had rather imputing the idiosyncrasies not
to the spherical symmetry, but to the chosen indicator of singularity,
and whether a more reliable indicator might exist. At variance with the
polynomial scalars built with the Riemann tensor, in the case
of the gamma metric this indicator, hopefully endowed with physical meaning,
should not exhibit a discontinuous
behavior, as a function of $k$, when $\rho\rightarrow 2l$.
In the search of such an alternative indicator,
we can look into the history of general relativity, and draw
inspiration from the ideas about the entity representing the gravitational
field that prevailed before Synge set his identification \cite{Synge1966}
of the gravitational field with the curvature of spacetime. For Synge that
assertion was not an abstract, a priori statement; it was
instead rooted in the sort of physical reasoning that had led him,
already in 1937, to reject \cite{Synge1937} Whittaker's
identification of the gravitational pull \cite{Whittaker1935}.\par
In compliance with Einstein's principle of equivalence
\cite{Einstein1916}, Whittaker had assumed that the gravito-inertial
force exerted on a pole test particle of unit mass, whose worldline was
a congruence with unit tangent vector $u^i(x^k)$, was defined by
minus the first curvature vector $a^i$ of the congruence:
\begin{equation}\label{15}
a^i=\frac{\dd u^i}{\dd s}+\Gamma^i_{kl}u^ku^l,
\end{equation}
namely, by minus the four-acceleration of the test particle.
If one thinks that in general relativity the gravito-inertial pull
should balance the nongravitational forces, (\ref{15}) can be read
like the extension of Newton's second law to the new theory.
This extension does not need to be postulated:
it can be derived from the conservation identities of the
theory\footnote{For the case of an
electrically charged pole test particle see the derivation by
Papapetrou and Urich \cite{PU1955}.},
and it indeed asserts that the force per unit mass $a^i$ exerted
by, say,  the electromagnetic field on the pole test particle is
balanced by minus the gravito-inertial pull per unit mass expressed
by the right-hand side of (\ref{15}).\par
Synge criticised Whittaker's definition of the gravito-inertial
pull on the ground that in
Einstein's theory only relative kinematic measurements
involving nearby particles are permitted in general, hence one
must renounce the unattainable goal of determining absolutely the force
acted by the gravitational field on any particle, and must be content
with a differential law that only allows for the comparison of the
gravitational pull acting at adjacent events.
Let us consider then two pole test particles, both executing geodesic motion, and
imagine that their world-lines $L$, $M$ be very close to each other.
If $\eta^i$ is the infinitesimal displacement vector drawn perpendicular
to $L$ from a point $A$ on $L$ to a point $B$ on $M$, the acceleration
of $B$  relative to $A$ is defined by the infinitesimal vector
\begin{equation}\label{16}
f^i=\frac{\DD^2\eta^i}{\dd s^2},
\end{equation}
where $\DD/\dd s$ indicates absolute differentiation and $\dd s$
is the infinitesimal arc length of the geodesic $L$ measured at $A$.
But Synge himself \cite{Synge1934} had proved that
\begin{equation}\label{17}
\frac{\DD^2\eta^i}{\dd s^2}+R^i_{~jkl}u^j\eta^ku^l=0,
\end{equation}
where $R^i_{~jkl}$ is the Riemann tensor and $u^i$ is the
four-velocity of the particle at $A$. By appealing to Newton,
Synge postulated that the excess of the gravitational force
at the event $B$ over the gravitational force at the event $A$ is
defined (for unit test masses) to be the acceleration (\ref{16})
of $B$ relative to $A$. Hence he found \cite{Synge1937} that
the excess of the gravitational pull is given by
\begin{equation}\label{18}
f^i=-R^i_{~jkl}u^j\eta^ku^l.
\end{equation}
Therefore, leaving behind Einstein's principle of equivalence,
through the equation of geodesic deviation
the identification between gravitational field and
curvature entered the theory of general relativity;
subsequently the polynomial invariants built with the Riemann tensor
became indicators of the singularity and \textit{ ad ignorantiam}
regularity of the gravitational field.\par
One may observe, however, that just in the case of the
static manifolds Synge's objection, that in Einstein's
theory only relative kinematic measurements are permitted,
hence ``one must be content with a differential law that only
allows for the comparison of the gravitational pull acting
at adjacent events'', is not true.\par
To understand why, it is necessary a sharpening of the definition
of static manifold, that is generally overlooked. We have reported
earlier that definition, both in the coordinate form and in the
intrinsic one, provided by equations (\ref{1}), according to which
a manifold is static if it is endowed with a timelike, hypersurface
orthogonal Killing vector. Such a definition, however, applies
also to the Minkowski spacetime.\par
The attribution of the adjective ``static'' to the Minkowski
spacetime does not seem however to be an appropriate one.
In fact, the notion of staticity is undisputably associated with
the notion of rest. How is it possible to call ``static'' the metric
of the theory of special relativity, a theory that
denies intrinsic meaning to the very notion of rest? How is it
possible that the Minkowski metric remain static, as it does
according to our definition, after we have subjected it to an
arbitrary Lorentz transformation, i.e. a transformation that
entails relative, uniform motion? Let us start from the Minkowski
metric, given by $g_{ik}={\rm diag}(-1,-1,-1,1)$ with respect
to the Galilean coordinates $x$, $y$, $z$, $t$, and perform the
coordinate transformation
\begin{equation}\label{19}
x=X\cosh{T}, \ y=Y, \ z=Z, \ t=X\sinh{T}
\end{equation}
to new coordinates $X$, $Y$, $Z$, $T$.
We get the particular Rindler metric \cite{Levi-Civita1918},
whose interval reads
\begin{equation}\label{20}
\dd s^2=-\dd X^2-\dd Y^2-\dd Z^2+X^2\dd T^2.
\end{equation}
How is it possible that this metric turn out to be in static form too,
despite the fact that the transformation
(\ref{19}) entails uniformly accelerated motion?
The definition of static manifold needs to be completed by
further specifying that the timelike, hypersurface orthogonal
Killing vector must be \textit{ uniquely} defined by equations (\ref{1})
at each event. This is the case, for instance, with such solutions
of general relativity as just Schwarzschild's original solution
and the gamma metric, provided that the curvature is nonvanishing.
In these manifolds of general relativity the congruences built
with the unique timelike, hypersurface orthogonal Killing vector
field define wordlines of absolute rest. They are intrinsic to the
manifold since, at variance with the worldlines of observers
arbitrarily drawn on a manifold, they are defined by the metric alone
through equations (\ref{1}). In this case it is by no means true that only
relative kinematic measurements are allowed, because a single such worldline
can be recognised in an absolute way, from the local
knowledge of the metric and of its derivatives.
From here it follows that the first curvature
(\ref{15}) of the worldlines of absolute rest is a local and intrinsic
property of a truly static manifold. There is no valid objection
to Whittaker's definition of the gravitational pull in this case:
the norm $a=(-a_ia^i)^{1/2}$ in this case is an
invariant, local, intrinsic quantity, defined only by the
metric, just like the polynomial invariants built with the Riemann
tensor happen to be. In the case of the gamma metric, defined with the spheroidal
coordinates  (\ref{10}) by equations (\ref{11}) and (\ref{12}), the
squared norm of the four-acceleration of a test particle on
a worldline of absolute rest reads \cite{ALM2003}
\begin{equation}\label{21}
a^2=\frac{16k^2l^2}{[4(\varrho-l)^2-4l^2\cos^2\vartheta]^{1-k^2}
(2\varrho-4l)^{1-k+k^2}
(2\varrho)^{1+k+k^2}},
\end{equation}
Therefore, for any value of $k<2$ and for all $\vartheta$ the norm
$a$ happens to grow without limit as $\varrho\rightarrow 2l$.
At variance with what occurs with the Kretschmann scalar, for
$\varrho\rightarrow 2l$, $a$ does not exhibit any discontinuous
behaviour when crossing the value $k=1$, for which
the gamma metric specialises into the Schwarzschild metric,
and the norm reduces to
\begin{equation}\label{22}
a=\left[\frac{l^2}{\varrho^3(\varrho-2l)}\right]^{1/2}.
\end{equation}
The invariant, local, intrinsic quantity $a$, if used as
indicator of singularity of a static manifold, does not
suffer from the ``idiosyncrasies'' exhibited by the
Kretschmann scalar $K$ of equation (\ref{13}): $a$ uniformly diverges when
$\varrho\rightarrow 2l$, thereby signalling
the presence of an invariant, local, intrinsic singularity
just on the inner border of Schwarzschild's original manifold,
hence of an inadmissible singularity in the interior of Hilbert's manifold.
\section{The singularity at the Schwarzschild surface is both intrinsic
and physical}
Like Synge's definition (\ref{18}) of the relative gravitational pull, Whittaker's
definition of the gravitational force felt by a test body of unit mass kept
at rest in a static field is connected to the geometric definition of the
corresponding four-acceleration (\ref{15}) by way of
hypothesis. It is true that the hypothesis
is supported by the derivation from the Einstein-Maxwell equations
done by Papapetrou and Urich \cite{PU1955}.
However, given the crucial r\^ole that Whittaker's definition has in
determining the intrinsic singular character of Schwarzschild's
surface, it seems worth calculating this force directly from some
actually existing exact solution of Einstein's equations. In this
way one may hope to ascertain whether the divergence of
an invariant, local, intrinsic geometric entity is accompanied by
the divergence of a physical quantity.
In this section the norm of the four-force exerted on a test body
in Schwarzschild's field is obtained \cite{ALM2001} by
starting, in the footsteps of Weyl
\cite{Weyl1919},\cite{BW1922}, from a particular two-body
Weyl-Levi Civita solution of Einstein's equations, calculated in 1922 by R. Bach
\cite{BW1922}.\par
\begin{figure}[ht]
\includegraphics{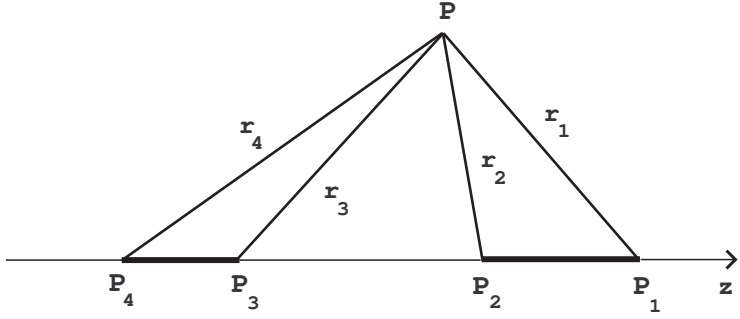}
\caption{\p Representation in the canonical $z$, $r$ half-plane of
the mass sources for Bach's two-body solution.
$r_4$, $r_3$ and $r_2$, $r_1$ are the ``distances'', calculated in
the Euclidean way, of a point $P$ from
the end points of the two segments endowed with mass.
$\overline{P_4P_3}=2l'$, $\overline{P_3P_2}=2d$, $\overline{P_2P_1}=2l$,
again in coordinate lengths.}
\end{figure}
By availing of  Weyl's canonical cylindrical coordinates,
this static, axially symmetric two-body solution is obtained if one assumes,
like Bach did, that the ``Newtonian potential'' $\psi$ entering
the line element (\ref{3})
is generated, in the canonical ``Bildraum",
by matter that is present with constant linear
mass density $\sigma=1/2$ on two segments of the symmetry axis, like the
segments $P_4P_3$ and $P_2P_1$ of Figure~3.
We know already from (\ref{7}) that the particular choice
\begin{equation}\label{23}
\psi=\frac{1}{2}\ln\frac{r_1+r_2-2l}{r_1+r_2+2l}
+\frac{1}{2}\ln\frac{r_3+r_4-2l'}{r_3+r_4+2l'},
\end{equation}
will produce a vacuum solution to Einstein's field equations
that reduces to Schwarz\-schild's original solution if one sets
either $l=0$ or $l'=0$. Of course, due to the nonlinearity of
(\ref{5}), one cannot expect that $\gamma$ will contain only the
sum of the contributions
\begin{equation}\label{24}
\gamma_{11}=\frac{1}{2}
\ln\frac{(r_1+r_2)^2-4l^2}{4r_1r_2},\;
\gamma_{22}=\frac{1}{2}
\ln\frac{(r_3+r_4)^2-4{l'}^2}{4r_3r_4},
\end{equation}
corresponding to the individual terms of the
potential (\ref{23}); a further term is present, that Bach
called $\gamma_{12}$, and reads
\begin{equation}\label{25}
\gamma_{12}
=\ln\frac{lr_4-(l'+d)r_1-(l+l'+d)r_2}{lr_3-dr_1-(l+d)r_2}+c,
\end{equation}
where $c$ is a constant.
Since $\gamma$ must vanish at the spatial infinity, it must be
$c=\ln[d/(l+l')]$. With this choice of the constant
one eventually finds \cite{BW1922} that the line element of the two-body
solution is defined by the functions
\begin{eqnarray}\nonumber
e^{2\psi}=\frac{r_1+r_2-2l}{r_1+r_2+2l}
\cdot\frac{r_3+r_4-2l'}{r_3+r_4+2l'},\\\nonumber
e^{2\gamma}=\frac{(r_1+r_2)^2-4l^2}{4r_1r_2}
\cdot\frac{(r_3+r_4)^2-4{l'}^2}{4r_3r_4}\\\label{26}
\cdot\left(\frac{d(l'+d)r_1+d(l+l'+d)r_2-ldr_4}
{d(l'+d)r_1+(l+d)(l'+d)r_2-l(l'+d)r_3}\right)^2.
\end{eqnarray}
With these definitions for $\psi$ and $\gamma$ the line
element (\ref{3}) behaves properly at the spatial infinity
and is regular everywhere, except for the two segments $P_4P_3$,
$P_2P_1$ of the symmetry axis, where the sources of $\psi$
are located, and also for the segment $P_3P_2$, because there
$\gamma$ does not vanish as required, but takes the constant
value
\begin{equation}\label{27}
\Gamma=\ln{\frac{d(l+l'+d)}{(l+d)(l'+d)}},
\end{equation}
thus giving rise to the well known conical singularity.\par

Due to this lack of elementary flatness occurring on the segment
$P_3P_2$, the solution is not a true two-body solution;
nevertheless Weyl showed \cite{BW1922} that a regular solution
could be obtained from it, provided that nonvanishing energy
tensor density $\mathbf{T}^k_i$ be allowed for in the space
between the two bodies. In this way an axial force $F$ is introduced,
with the evident function of keeping the two bodies at rest
despite their mutual gravitational attraction.
By providing a measure for $F$, Weyl provided a measure of the
gravitational pull. Let us recall here Weyl's analysis
\cite{Weyl1919, BW1922} of the axially symmetric, static
two-body problem.\par
In writing Einstein's field equations, we adopt henceforth Weyl's
convention for the energy tensor:
\begin{equation}\label{28}
R_{ik}-\frac{1}{2}g_{ik}R=-T_{ik}.
\end{equation}
Einstein's equations teach that, when the line element has the
expression (\ref{3}), $\mathbf{T}^k_i$ shall have the form
\begin{equation}\label{29}
\left(\begin{array}{llll}
\mathbf{T}^0_0&0&0&0\\
\\
0&\mathbf{T}^1_1&\mathbf{T}^1_2&0\\
\\
0&\mathbf{T}^2_1&\mathbf{T}^2_2&0\\
\\
0&0&0&\mathbf{T}^3_3
\end{array}\right)
\end{equation}
where
\begin{equation}\label{30}
\mathbf{T}^1_1+\mathbf{T}^2_2=0.
\end{equation}
By introducing the notation
\begin{equation}\label{31}
\mathbf{T}^3_3=r\varrho',\;\mathbf{T}^0_0=r(\varrho+\varrho'),
\end{equation}
Einstein's equations can be written as:
\begin{equation}\label{32}
\Delta\psi=\frac{1}{2}\varrho,\;
\frac{\partial^2\gamma}{\partial z^2}
+\frac{\partial^2\gamma}{\partial r^2}
+\left\{\left(\frac{\partial\psi}{\partial z}\right)^2
+\left(\frac{\partial\psi}{\partial r}\right)^2\right\}
=-\varrho';
\end{equation}
\begin{equation}\label{33}
\mathbf{T}^1_1=-\mathbf{T}^2_2
=\gamma_r-r(\psi^2_r-\psi^2_z),\;
-\mathbf{T}^2_1=-\mathbf{T}^1_2=\gamma_z-2r\psi_r\psi_z.
\end{equation}
Weyl shows that $\varrho$ must be interpreted as mass density in
the canonical ``Bildraum''.
To this end he considers the mass density
distribution sketched in Figure 4, where $\varrho$ is assumed to be
nonvanishing only in the shaded regions labeled $1$ and $2$ respectively.
\begin{figure}[ht]
\includegraphics{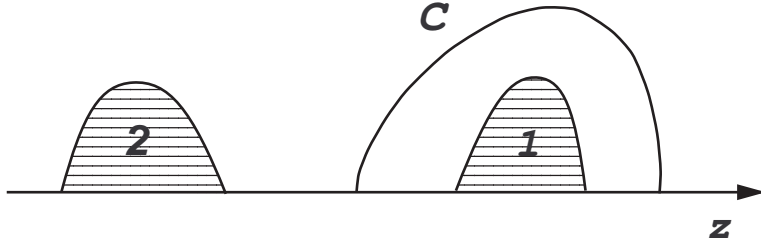}
\caption{\p Representation in the canonical $z$, $r$ half-plane
of extended mass sources of a two-body solution.}
\end{figure}
According to (\ref{23}), the potential $\psi$ corresponding to this
mass distribution can be uniquely split in two terms $\psi_1$ and
$\psi_2$, such that $\psi_1$ is a potential function that vanishes
at infinity and is everywhere regular outside the region $1$,
while $\psi_2$ behaves in the same way outside the region $2$. The
asymptotic forms of $\psi_1$ and $\psi_2$ are such that
\begin{equation}\label{34}
e^{2\psi_1}=1-\frac{m_1}{R}+\cdot\cdot\cdot,\;
e^{2\psi_2}=1-\frac{m_2}{R}+\cdot\cdot\cdot
\end{equation}
where the mass coefficients $m_1$ and $m_2$ are given by the
integral $\int{\varrho \dd V}=2\pi\int{\varrho r\dd r\dd z}$, performed
in the canonical space and extended to the appropriate shaded
region. Outside the shaded regions one has $\varrho=0$, but
there shall be some region between the bodies, let us call it $L'$,
where $\varrho=0$ but $\mathbf{T}^k_i\neq 0$, since in a static
solution of general relativity the gravitational pull shall be
counteracted in some way. Weyl's procedure for determining
$\mathbf{T}^k_i$ in $L'$ is the following. Suppose that $\mathbf{T}^k_i$
vanishes outside a simply connected region $L$ that includes both material bodies.
Since $\psi$ is known there, we can avail of (\ref{6}), together with
the injunction that $\gamma$ vanish at infinity, to determine $\gamma$
uniquely outside $L$. Within $L'$ we can choose $\gamma$
arbitrarily, provided that we ensure the regular connection with
the vacuum region and the regular behaviour on the axis,
i.e. $\gamma$ vanishing there like $r^2$. Since $\psi$ is
known in $L'$ and $\gamma$ has been chosen as just shown, we can
use equations (\ref{32}) and (\ref{33}) to determine
$\mathbf{T}^k_i$ there.\par
If the material bodies include each one a segment of the axis, just as it
occurs in Fig. 4, the force $F$ directed along the $z$ axis,
with which the stresses in $L'$ contrast the gravitational pull
can be written as
\begin{equation}\label{35}
F=2\pi\int_C(\mathbf{T}^2_1 \dd z-\mathbf{T}^1_1 \dd r);
\end{equation}
the integration path is along a curve $C$, like the one drawn in
Fig. 4, that separates the two bodies in the meridian half-plane;
the value of the integral does not depend on the precise position
of $C$ because, as one gathers from the definitions (\ref{32}),
(\ref{33}):
\begin{equation}\label{36}
\mathbf{T}^1_{1,1}+\mathbf{T}^2_{1,2}=0
\end{equation}
in the region $L'$. Since the region of the meridian half-plane
where $\varrho=0$ is simply connected, by starting from $\psi$
and from the vacuum equation (\ref{6}), now rewritten as:
\begin{equation}\label{37}
\dd\gamma^*=2r\psi_z\psi_r\dd z+r(\psi^2_r-\psi^2_z)\dd r
\end{equation}
one can uniquely define there the function $\gamma^*$ that
vanishes at the spatial infinity. In all the
parts of the $z$ axis where $\varrho=0$ it must be $\gamma^*_z=0$,
$\gamma^*_r=0$, hence $\gamma^*=$const., $\gamma^*_r=0$. In
particular, in the parts of the axis that go to infinity one shall
have $\gamma^*=0$; let us call $\Gamma^*$ the constant value
assumed instead by $\gamma^*$ on the segment of the axis lying
between the two bodies. The definitions (\ref{33}) can now be
rewritten as:
\begin{equation}\label{38}
\mathbf{T}^1_1=-\mathbf{T}^2_2=\gamma_r-\gamma^*_r,\;
-\mathbf{T}^2_1=-\mathbf{T}^1_2=\gamma_z-\gamma^*_z,
\end{equation}
and the integral of (\ref{35}) becomes
\begin{equation}\label{39}
\int_C(\mathbf{T}^2_1 \dd z-\mathbf{T}^1_1 \dd r)
=\int_C{(\gamma^*_z-\gamma_z)\dd z+(\gamma^*_r-\gamma_r)\dd r}
=\int_C \dd(\gamma^*-\gamma).
\end{equation}
Since $\gamma$ vanishes on the parts of the $z$ axis where
$\varrho=0$, the force $F$ that holds the bodies at rest
despite the gravitational pull shall be
\begin{equation}\label{40}
F=-2\pi\Gamma^*
\end{equation}
with Weyl's definition (\ref{28}) of the energy tensor. When the
mass density $\varrho$ has in the canonical space the particular
distribution considered by Bach and drawn in Fig. 3,
$\Gamma^*$ is equal to $\Gamma$ as defined by (\ref{27}).
The measure of the gravitational pull with which the two ``material
bodies'' of this particular solution attract each other therefore turns
out to be
\begin{equation}\label{41}
F=2\pi\ln{\frac{(d+l)(d+l')}{d(d+l+l')}}
\end{equation}
in Weyl's units. This expression agrees with the Newtonian value
when $l$ and $l'$ are small when compared to $d$, as expected.\par

Despite its mathematical beauty, Weyl's definition of the
gravitational pull for an axially symmetric, static two-body
solution appears associated without remedy to the adoption
of the canonical coordinate system. It is however possible to
obtain through Weyl's definition of $F$, given by (\ref{35}),
a ``quasi'' four-vector $f_i$. In fact that expression can be
rewritten as
\begin{equation}\label{42}
F=\int_{\Sigma}\mathbf{T}^l_1 \dd f^*_{0l}
\equiv\frac{1}{2}\int_{\Sigma} T^l_1\epsilon_{0lmn}\dd f^{mn},
\end{equation}
where $\epsilon_{klmn}$ is Levi-Civita's totally antisymmetric
tensor and $\dd f^{mn}$ is the element of the two-surface $\Sigma$
generated by the curve $C$ through rotation around the symmetry
axis. Since the metric that we are considering
is static in the strict sense it is possible
to define a unique timelike, hypersurface orthogonal Killing
vector $\xi^k_{(t)}$ that correspond, in Weyl's
canonical coordinates, to a unit coordinate time translation. Therefore
(\ref{42}) can be rewritten as
\begin{equation}\label{43}
F=\frac{1}{2}\int_{\Sigma}\xi^k_{(t)}T^l_1\epsilon_{klmn}\dd f^{mn}
\end{equation}
by still using the canonical coordinates. Now the integrand is
written as the first component of the infinitesimal
covariant four-vector
\begin{equation}\label{44}
\xi^k_{(t)}T^l_i\epsilon_{klmn}\dd f^{mn},
\end{equation}
but of course in general the expression
\begin{equation}\label{45}
f_i=\frac{1}{2}\int_{\Sigma}
\xi^k_{(t)}T^l_i\epsilon_{klmn}\dd f^{mn}
\end{equation}
will not be a four-vector, because the integration over $\Sigma$ spoils
the covariance. When evaluated in canonical coordinates, the
nonvanishing components of $f_i$ are $f_1=F$ and
\begin{equation}\label{46}
f_2=2\pi\int_C(\mathbf{T}^2_2 \dd z-\mathbf{T}^2_1 \dd r)
=2\pi\int_C{(\gamma^*_r-\gamma_r)\dd z-(\gamma^*_z-\gamma_z)\dd r},
\end{equation}
that however must vanish too, if $f_i$ has to become a four-vector
defined on the symmetry axis. But, as one sees from
Weyl's analysis, we are at freedom to choose $\mathbf{T}^k_i$ in
$L'$ as nonvanishing only in a tube with a very small, yet finite coordinate
radius, that encloses in its interior the segment of the symmetry
axis lying between the bodies; moreover, we can freely set
$\gamma_z=\gamma^*_z$ within the tube. Under these conditions
the second term of the integral (\ref{46}) just vanishes, while
the first one shall be very small, since the regularity of the surface
$\Sigma$ requires that the curve $C$ approach the symmetry axis at a
right angle in canonical coordinates. By properly choosing
$\mathbf{T}^k_i$ we thus succeed in providing through equation
(\ref{45}) a quasi four-vector $f_i$ whose components, written
in Weyl's canonical coordinates, reduce in approximation to
$(F,0,0,0)$.\par
Having defined, with the above caveats, the quasi four-vector $f_i$
along the segment of the symmetry axis between the two bodies,
we can use its ``quasi'' norm to provide a measure of the force
that opposes the gravitational pull. In the case of Bach's two-body
solution, whose line element is defined in canonical coordinates
by (\ref{3}) and (\ref{26}), that quasi norm reads
\begin{equation}\label{47}
f\equiv(-f^if_i)^{1/2}
=2\pi\ln\frac{(d+l)(d+l')}{d(d+l+l')}
\cdot\left[\frac{r_1-2l}{r_1}
\cdot\frac{r_4-2l'}{r_4}\right]^{1/2}
\end{equation}
when measured in Weyl's units at a point of the symmetry
axis for which $z_3<z<z_2$. At variance with the behaviour of $F$,
the quasi norm $f$ depends on $z$, due to the term of (\ref{47})
enclosed within the square brackets, that comes from $e^{2\psi}$.
Let us evaluate this quasi norm divided by $l'$ when $l'\rightarrow 0$,
namely, the coefficient of the linear term in the McLaurin
series expansion of $f$ with respect to $l'$. Since $\Gamma^*$,
now defined by the right-hand side of (\ref{27}),
tends to zero when $l'\rightarrow 0$, while performing this limit one can
also send to zero the radius of the very narrow tube considered in the
previous section. Therefore $f_i$ can become a true four-vector and $f$
can become a true norm in the above mentioned limit. With this
proviso one finds the invariant, exact result
\begin{equation}\label{48}
\lim_{l'\rightarrow 0}\left[\frac{f}{l'}\right]
=\left[\frac{\partial f}{\partial l'}\right]_{l'=0}
=\frac{2\pi l}{d(d+l)}\left(\frac{r_1-2l}{r_1}\right)^{1/2}.
\end{equation}
When $l'\rightarrow 0$ the line element of Bach's solution
with two bodies tends to the line element defined by (\ref{3}) and
(\ref{7}) when $k=1$, that is in one-to-one correspondence with the
line element of Schwarzschild's original solution \cite{Schwarzschild1916}.
Therefore the scalar quantity $[\partial f/\partial l']_{l'=0}$
evaluated at $P_3$ shall be the norm of the force per unit mass
exerted by Schwarzschild's gravitational field on a test
particle kept at rest at $P_3$. Its value is obtained by substituting
$2d+2l$ for $r_1$ in (\ref{48}). One finds
\begin{equation}\label{49}
\left(\lim_{l'\rightarrow 0}\left[\frac{f}{l'}\right]\right)_{z=z_3}
=\frac{8\pi l}{(2d+2l)^{3/2}(2d)^{1/2}}.
\end{equation}
If one solves Schwarzschild's problem in spherical polar coordinates $r$,
$\vartheta$, $\varphi$, $t$ with three unknown functions $\lambda(r)$,
$\mu(r)$, $\nu(r)$, i.e. without fixing the radial coordinate, like
Combridge and Janne did long ago \cite{Combridge1923},\cite{Janne1923},
one ends up writing de~Sitter's line element \cite{de Sitter1916}
\begin{equation}\label{50}
\dd s^2=-\exp{\lambda}\dd r^2
-\exp{\mu}[r^2(\dd\vartheta^2+\sin^2{\vartheta}\dd\varphi^2)]
+\exp{\nu}\dd t^2
\end{equation}
in terms of one unknown function $h(r)$. In fact $\lambda$,
$\mu$, $\nu$ are defined through this arbitrary function $h(r)$
and through its derivative ${h'}(r)$ as follows:
\begin{eqnarray}\label{51}
\exp{\lambda}=\frac{{h'}^2}{1-2m/h},\\\label{52}
\exp{\mu}=\frac{h^2}{r^2},\\\label{53}
\exp{\nu}=1-2m/h.
\end{eqnarray}
Here $m$ is the mass constant; of course the arbitrary function
$h$ must have the appropriate
behaviour as $r\rightarrow +\infty$. Schwarzschild's original
solution \cite{Schwarzschild1916} is eventually recovered
\cite{Abrams1979},\cite{Abrams1989} by requiring that $h$ be a monotonic
function of $r$ and that $h(0)=2m$. With our symmetry-adapted coordinates,
the unique worldline of absolute rest of a test body shall be invariantly
specified by requiring that the spatial coordinates $r$, $\vartheta$, $\varphi$
of the test body be constant in time. If $a=(-a_ia^i)^{1/2}$
is the norm of the acceleration four-vector (\ref{15})
along the worldline of the test body, one finds
\begin{equation}\label{54}
a=\frac {m}{h^{3/2}(h-2m)^{1/2}},
\end{equation}
in keeping with the expression (\ref{22}), derived with a
particular choice of the radial coordinate.
This norm was postulated by Whittaker \cite{Whittaker1935} to be equal to
the norm of the force per unit mass needed for constraining the
test particle to follow a worldline of absolute rest despite the
gravitational pull of the Schwarzschild field.
The consistency of the hypothesis with Einstein's theory requires
that $a$ be equal to the scalar quantity
$[\partial f/\partial l']_{l'=0,\;z=z_3}$
that provides the norm of the force per unit mass for Bach's
solution in the test particle limit $l'\rightarrow 0$.\par
This is indeed the case, since the functional dependence of
(\ref{49}) on the mass parameter $l$ and on the coordinate distance
$2d+2l$ is the same as the functional dependence of (\ref{54})
on the mass parameter $m$ and on the function $h(r)$, for which
$h(0)=2m$, introduced above.
The extra constant $8\pi$ appearing in
(\ref{49}) is just due to Weyl's adoption of the definition (\ref{28})
of the energy tensor. For Schwarzschild's field, the
definition of the norm of the gravitational force exerted on a test particle at rest
obtained through the acceleration four-vector and the independent definition
through the force that, in Bach's two-body solution, $\mathbf{T}^k_i$
must exert to keep the masses at rest when $l'\rightarrow 0$ lead to
one and the same result. In particular, both definitions show
that the norm of the force per unit mass grows without limit as the
test particle is kept at rest in a position closer and closer to
Schwarzschild's two-surface. Therefore the singularity at the
Schwarzschild surface, besides being invariant, local and
intrinsic, is also physical in character.

\section{Conclusion}

Several reasons for reinstating Schwarzschild's original solution
and manifold \cite{Schwarzschild1916} as the idealised model for the field of a
``Massenpunkt'' in the theory of general relativity have been expounded in the
previous sections. The model happens to fall short of some of
our expectations about the field of a material particle. One still feels in
want of something as simple and essential as its Newtonian counterpart.
Confronted with the singular surface of finite area at the inner
border of the manifold, one may well ask, with Marcel Brillouin
\cite{Brillouin1923}, where the point particle has gone.
However, given the astounding achievements of just this model,
one cannot help following Brillouin in his resigned way of
accepting the Schwarzschild manifold\footnote{In the quoted paper he wrote:
\begin{quotation}
Toutefois, comme on ne peut rien trouver de plus ponctuel dans
l'Univers d'Einstein, et qu'il faut bien arriver a definir le
\textit{corps d'\'epreuve mat\'eriel} \'el\'ementaire qui,
d'apr\`es Einstein, suit une g\'eod\'esique de l'Univers dont il
fait partie, je conserverai cette \textit{d\'enomination
abr\'eg\'ee, point materiel}, sans oublier son imperfection.
\end{quotation}
}.
At last, one may add in consolation, unlike the Hilbert manifold, it is
endowed with a consistent arrow of time.
Moreover, unlike both the Hilbert manifold and its Kruskal-Szekeres
extension, it does not exhibit an invariant, local, intrinsic
singularity in its interior.\par
While studying Schwarzschild's problem, some circumstances of a more
general theoretical character have been reconsidered, that are worth being
severally recalled here:
\begin{itemize}
\item Einstein's equations do not generally fix the topology.
\item When we accept that the topology of a space-time has physical consequences,
providing a solution to Einstein's equations implies the statement of the topology,
and locally isometric solutions are different when their topologies are.
\item Locally isometric space-times change their topology when
singular coordinate transformations are applied.
\item The invariants of the Riemann tensor
indicate singularities. In the generic case, they are always involved.
In the algebraically special space-times, there
are instances where the Killing congruence may exhibit singularities without
involvement of the local Riemann tensor. The Schwarzschild horizon
is one example.
\end{itemize}

\appendix
\section{Translation of Schwarzschild's \\ ``Massenpunkt''
paper}\label{A}
\bigskip
\centerline{\gr On the Gravitational Field of a Mass Point
according to Einstein's Theory\footnote{Original title:
\"Uber das Gravitationsfeld eines Massenpunktes nach
der Einsteinschen Theorie. Published in: Sitzungsberichte der K\"oniglich
Preussischen Akademie der Wissenschaften zu Berlin, Phys.-Math. Klasse
1916, 189-196. Submitted January 13, 1916. Translation by S.
Antoci, Dipartimento di Fisica ``A. Volta'', Universit\`a di Pavia,
and A. Loinger, Dipartimento di Fisica, Universit\`a di Milano.
The valuable advice of D.-E. Liebscher is gratefully
acknowledged.}}\medskip
\centerline{\bf K. Schwarzschild}
\bigskip
% ----------------------------------------------------------------
\S 1. In his work on the motion of the perihelion of Mercury (see
Sitzungsberichte of November 18, 1915) Mr. Einstein has posed
the following problem:\par
Let a point move according to the prescription:
\begin{eqnarray}\nonumber
&\delta\int{\dd s}=0,\\\label{A1}
&{\rm where}\\\nonumber
&\dd s=\sqrt{\Sigma g_{\mu\nu}\dd x_\mu
\dd x_\nu} \ \ {\mu, \nu=1, 2, 3, 4},
\end{eqnarray}
where the $g_{\mu\nu}$ stand for functions of the variables $x$, and
in the variation the variables $x$ must be kept fixed at the
beginning and at the end of the path of integration. In short, the
point shall move along a geodesic line in the manifold
characterised by the line element $ds$.\par
The execution of the variation yields the equations of motion of the
point:
\begin{equation}\label{A2}
\frac{\dd ^2x_\alpha}{\dd s^2}={\sum_{\mu,\nu}}~\Gamma^\alpha_{\mu\nu}
\frac{\dd x_\mu}{\dd s}\frac{\dd x_\nu}{\dd s}, \ \ \alpha, \beta=1, 2, 3, 4,
\end{equation}
where
\begin{equation}\label{A3}
\Gamma^\alpha_{\mu\nu}=-\frac{1}{2}\sum_\beta g^{\alpha\beta}
\bigg(\frac{\partial g_{\mu\beta}}{\partial x_\nu}+
\frac{\partial g_{\nu\beta}}{\partial x_\mu}
-\frac{\partial g_{\mu\nu}}{\partial x_\beta}\bigg),
\end{equation}
and the $g^{\alpha\beta}$ stand for the normalised minors associated to
$g_{\alpha\beta}$ in the determinant $\vert g_{\mu\nu}\vert$.\par
According to Einstein's theory, this is the motion of a
massless point in the gravitational field of a mass at
the point $x_1=x_2=x_3=0$, if the ``components of the
gravitational field'' $\Gamma$ fulfill everywhere, with the
exception of the point $x_1=x_2=x_3=0$, the ``field equations''
\begin{equation}\label{A4}
\sum_\alpha\frac{\partial \Gamma^\alpha_{\mu\nu}}
{\partial x_\alpha}
+\sum_{\alpha\beta}~\Gamma^\alpha_{\mu\beta
}\Gamma^\beta_{\nu\alpha}=0,
\end{equation}
and if also the ``equation of the determinant''
\begin{equation}\label{A5}
\vert g_{\mu\nu}\vert=-1
\end{equation}
is satisfied.\par
The field equations together with the equation of the
determinant have the fundamental property that they preserve their
form under the substitution of other arbitrary variables in lieu
of $x_1$, $x_2$, $x_3$, $x_4$, as long as the determinant of the
substitution is equal to $1$.\par
Let $x_1$, $x_2$, $x_3$ stand for rectangular co-ordinates, $x_4$ for the
time; furthermore, the mass at the origin shall not change with time, and the
motion at infinity shall be rectilinear and uniform. Then,
according to Mr. Einstein's list, \textit{ loc. cit.} p. 833, the following
conditions must be fulfilled too:\smallskip
\begin{enumerate}
\item All the components are independent of the time $x_4$.
\item The equations $g_{\rho4}=g_{4\rho}=0$ hold exactly for
$\rho=1, 2, 3.$
\item The solution is spatially symmetric with respect to the
origin of the co-ordinate system in the sense that one finds again
the same solution when $x_1$, $x_2$, $x_3$ are subjected to an
orthogonal transformation (rotation).
\item The $g_{\mu\nu}$ vanish at infinity, with the exception
of the following four limit values different from zero:
$$
g_{44}=1,~~g_{11}=g_{22}=g_{33}=-1.
$$
\end{enumerate}
\par\noindent
\textit{ The problem is to find out a line element with
coefficients such that the field equations, the equation of the
determinant and these four requirements are satisfied.}\par
\S 2. Mr. Einstein showed that this problem, in first
approximation, leads to Newton's law and that the second
approximation correctly reproduces the known anomaly in the motion
of the perihelion of Mercury. The following calculation yields
the exact solution of the problem. It is always pleasant to avail
of exact solutions of simple form. More importantly, the
calculation proves also the uniqueness of the solution, about
which Mr. Einstein's treatment still left doubt, and which could have
been proved only with great difficulty, in the way shown below,
through such an approximation method. The following lines
therefore let Mr. Einstein's result shine with increased
clearness.\par
\S 3. If one calls $t$ the time, $x$, $y$, $z$ the rectangular
co-ordinates, the most general line element that satisfies the
conditions (1)-(3) is clearly the following:
$$
\dd s^2=F\dd t^2-G(\dd x^2+\dd y^2+\dd z^2)-H(x\dd x+y\dd y+z\dd z)^2
$$
where $F$, $G$, $H$ are functions of $r=\sqrt{x^2+y^2+z^2}$.\par
The condition (4) requires: for $r=\infty: F=G=1, H=0$.\par
When one goes over to polar co-ordinates according to
$x=r\sin\vartheta\cos\phi$, $y=r\sin\vartheta\sin\phi$,
$z=r\cos\vartheta$,
the same line element reads:
\begin{eqnarray}\label{A6}
\dd s^2&=F\dd t^2
-G(\dd r^2+r^2\dd \vartheta^2+r^2sin^2\vartheta
\dd \phi^2)-Hr^2\dd r^2\\\nonumber
&=F\dd t^2-(G+Hr^2)\dd r^2-Gr^2(\dd \vartheta^2+sin^2\vartheta
\dd \phi^2).
\end{eqnarray}

Now the volume element in polar co-ordinates is equal to
$r^2\sin\vartheta \dd r\dd\vartheta\dd\phi$, the functional determinant
$r^2\sin\vartheta$ of the old with respect to the new coordinates
is different from $1$; then the field equations would not remain in
unaltered form if one would calculate with these polar
co-ordinates, and one would have to perform a cumbersome transformation.
However there is an easy trick to circumvent this difficulty. One
puts:
\begin{equation}\label{A7}
x_1=\frac{r^3}{3},~~x_2=-\cos\vartheta,~~x_3=\phi.
\end{equation}
Then we have for the volume element:
$r^2\dd r\sin\vartheta \dd\vartheta \dd\phi=\dd x_1\dd x_2\dd x_3.$ The new
variables are then \textit{ polar co-ordinates with the determinant 1}.
They have the evident advantages of polar co-ordinates for the
treatment of the problem, and at the same time, when
one includes also $t=x_4$, the field equations and the determinant
equation remain in unaltered form.\par
In the new polar co-ordinates the line element reads:
\begin{equation}\label{A8}
\dd s^2=F\dd x_4^2-\bigg(\frac{G}{r^4}+\frac{H}{r^2}\bigg)\dd x_1^2
-Gr^2\bigg[\frac{\dd x_2^2}{1-x_2^2}+\dd x_3^2(1-x_2^2)\bigg],
\end{equation}
for which we write:
\begin{equation}\label{A9}
\dd s^2=f_4\dd x_4^2-f_1\dd x_1^2-f_2\frac{\dd x_2^2}{1-x_2^2}
-f_3\dd x_3^2(1-x_2^2).
\end{equation}
Then $f_1$, $f_2=f_3$, $f_4$ are three functions of $x_1$ which
have to fulfill the following conditions:\par\smallskip
\begin{enumerate}
\item For $x_1=\infty:$ $f_1=\frac{1}{r^4}=(3x_1)^{-4/3}$,
$f_2=f_3=r^2=(3x_1)^{2/3}$, $f_4=1$.
\item The equation of the determinant:
$f_1\cdot f_2\cdot f_3\cdot f_4=1$.
\item The field equations.
\item Continuity of the $f$, except for $x_1=0$.
\end{enumerate}\par\smallskip
\S 4. In order to formulate the field equations one must first
form the components of the gravitational field corresponding to the
line element (\ref{A9}). This happens in the simplest way when one builds
the differential equations of the geodesic line by direct
execution of the variation, and reads out the components from these.
The differential equations of the geodesic line for the line
element (\ref{A9}) result from the variation immediately in the form:
\begin{eqnarray}\nonumber
0=f_1\frac{\dd ^2x_1}{\dd s^2}+\frac{1}{2}
\frac{\partial f_4}{\partial x_1}\bigg(\frac{\dd x_4}{\dd s}\bigg)^2
+\frac{1}{2}\frac{\partial f_1}{\partial x_1}
\bigg(\frac{\dd x_1}{\dd s}\bigg)^2\\\nonumber
-\frac{1}{2}\frac{\partial f_2}{\partial x_1}
\bigg[\frac{1}{1-x_2^2}\bigg(\frac{\dd x_2}{\dd s}\bigg)^2
+(1-x_2^2)\bigg(\frac{\dd x_3}{\dd s}\bigg)^2\bigg],\\\nonumber\\\nonumber
0=\frac{f_2}{1-x_2^2}\frac{\dd ^2x_2}{\dd s^2}+
\frac{\partial f_2}{\partial x_1}\frac{1}{1-x_2^2}
\frac{\dd x_1}{\dd s}\frac{\dd x_2}{\dd s}\\\nonumber
+\frac{f_2x_2}{(1-x_2^2)^2}\bigg(\frac{\dd x_2}{\dd s}\bigg)^2
+f_2x_2\bigg(\frac{\dd x_3}{\dd s}\bigg)^2,\\\nonumber\\\nonumber
0=f_2(1-x_2^2)\frac{\dd ^2x_3}{\dd s^2}+
\frac{\partial f_2}{\partial x_1}
(1-x_2^2)\frac{\dd x_1}{\dd s}\frac{\dd x_3}{\dd s}
-2f_2x_2\frac{\dd x_2}{\dd s}\frac{\dd x_3}{\dd s},\\\nonumber\\\nonumber
0=f_4\frac{\dd ^2x_4}{\dd s^2}+
\frac{\partial f_4}{\partial x_1}\frac{\dd x_1}{\dd s}\frac{\dd x_4}{\dd s}.
\end{eqnarray}

The comparison with (\ref{A2}) gives the components of the gravitational
field:
\begin{eqnarray}\nonumber
\Gamma^1_{11}=-\frac{1}{2}\frac{1}{f_1}
\frac{\partial f_1}{\partial x_1},~~~
\Gamma^1_{22}=+\frac{1}{2}\frac{1}{f_1}
\frac{\partial f_2}{\partial x_1}
\frac{1}{1-x_2^2},\\\nonumber
\Gamma^1_{33}=+\frac{1}{2}\frac{1}{f_1}
\frac{\partial f_2}{\partial x_1}(1-x_2^2),~~~
\Gamma^1_{44}=-\frac{1}{2}\frac{1}{f_1}
\frac{\partial f_4}{\partial x_1},\\\nonumber
\Gamma^2_{21}=-\frac{1}{2}\frac{1}{f_2}
\frac{\partial f_2}{\partial x_1},~~~
\Gamma^2_{22}=-\frac{x_2}{1-x_2^2},~~~
\Gamma^2_{33}=-x_2(1-x_2^2),\\\nonumber
\Gamma^3_{31}=-\frac{1}{2}\frac{1}{f_2}
\frac{\partial f_2}{\partial x_1},~~~
\Gamma^3_{32}=+\frac{x_2}{1-x_2^2},\\\nonumber
\Gamma^4_{41}=-\frac{1}{2}\frac{1}{f_4}
\frac{\partial f_4}{\partial x_1}\\\nonumber
{\rm(the~remaining~ones~are~zero).}
\end{eqnarray}

Due to the rotational symmetry
around the origin  it is sufficient to write the field equations
only for the equator ($x_2=0$); therefore, since they will be
differentiated only once, in the previous expressions it is
possible to set everywhere since the beginning $1-x_2^2$ equal
to $1$. The calculation of the field equations then gives
\begin{eqnarray}\nonumber
a)~\frac{\partial}{\partial x_1}\bigg(\frac{1}{f_1}
\frac{\partial f_1}{\partial x_1}\bigg)
=\frac{1}{2}\bigg(\frac{1}{f_1} \frac{\partial f_1}{\partial x_1}\bigg)^2
+\bigg(\frac{1}{f_2}
\frac{\partial f_2}{\partial x_1}\bigg)^2
+\frac{1}{2}\bigg(\frac{1}{f_4}
\frac{\partial f_4}{\partial x_1}\bigg)^2,\\\nonumber
b)~\frac{\partial}{\partial x_1}\bigg(\frac{1}{f_1}
\frac{\partial f_2}{\partial x_1}\bigg)
=2+\frac{1}{f_1f_2}\bigg( \frac{\partial f_2}{\partial x_1}\bigg)^2,\\\nonumber
c)~\frac{\partial}{\partial x_1}
\bigg(\frac{1}{f_1} \frac{\partial f_4}{\partial x_1}\bigg)
=\frac{1}{f_1f_4}\bigg( \frac{\partial f_4}{\partial x_1}\bigg)^2.
\end{eqnarray}
Besides these three equations the functions
$f_1$, $f_2$, $f_4$ must fulfill also the equation of the
determinant
$$
d)~f_1f_2^2f_4=1, \ \ {\rm or:} \ \ \frac{1}{f_1} \frac{\partial f_1}{\partial x_1}
+\frac{2}{f_2}\frac{\partial f_2}{\partial x_1}
+\frac{1}{f_4} \frac{\partial f_4}{\partial x_1}=0.
$$
For now I neglect ($b$) and determine the three functions $f_1$,
$f_2$, $f_4$ from ($a$), ($c$), and ($d$). ($c$) can be transposed into the
form

$$c')~\frac{\partial}{\partial x_1}\bigg(\frac{1}{f_4}
\frac{\partial f_4}{\partial x_1}\bigg) =\frac{1}{f_1f_4}
\frac{\partial f_1}{\partial x_1}
\frac{\partial f_4}{\partial x_1}.$$
This can be directly integrated and gives
$$
c'')~\frac{1}{f_4} \frac{\partial f_4}{\partial x_1} =\alpha
f_1, \ \ \ {\rm(\alpha~integration~constant)}
$$
the addition of ($a$) and ($c'$) gives
$$
\frac{\partial}{\partial x_1}\bigg(\frac{1}{f_1}
\frac{\partial f_1}{\partial x_1} +\frac{1}{f_4}
\frac{\partial f_4}{\partial x_1}\bigg)
=\bigg(\frac{1}{f_2} \frac{\partial f_2}{\partial x_1}\bigg)^2
+\frac{1}{2}\bigg(\frac{1}{f_1}
\frac{\partial f_1}{\partial x_1} +\frac{1}{f_4}
\frac{\partial f_4}{\partial x_1}\bigg)^2.
$$
By taking ($d$) into account it follows
$$
-2\frac{\partial}{\partial x_1}\bigg(\frac{1}{f_2}
\frac{\partial f_2}{\partial x_1}\bigg) =3\bigg(\frac{1}{f_2}
\frac{\partial f_2}{\partial x_1}\bigg)^2.
$$
By integrating
$$
\frac{1}{\frac{1}{f_2}\frac{\partial f_2}{\partial x_1}}=
\frac{3}{2}x_1+\frac{\rho}{2} \ \ \
{\rm(\rho~integration~constant)}
$$
or
$$
\frac{1}{f_2}\frac{\partial f_2}{\partial x_1}=\frac{2}{3x_1+\rho}.
$$
By integrating once more,
$$
f_2=\lambda(3x_1+\rho)^{2/3}. \ \ \
{\rm(\lambda~integration~constant)}
$$
The condition at infinity requires:
$\lambda=1$. Then
\begin{equation}\label{A10}
f_2=(3x_1+\rho)^{2/3}.
\end{equation}
Hence it results further from ($c''$) and ($d$)
$$
\frac{\partial f_4}{\partial x_1}=\alpha f_1f_4
=\frac{\alpha}{f_2^2}=\frac{\alpha}{(3x_1+\rho)^{4/3}}.
$$
By integrating while taking into account the condition at infinity
\begin{equation}\label{A11}
f_4=1-\alpha(3x_1+\rho)^{-1/3}.
\end{equation}
Hence from ($d$)
\begin{equation}\label{A12}
f_1=\frac{(3x_1+\rho)^{-4/3}}
{1-\alpha(3x_1+\rho)^{-1/3}}.
\end{equation}
As can be easily verified, the equation ($b$) is automatically
fulfilled by the expressions that we found for $f_1$ and $f_2$.\par
Therefore all the conditions are satisfied apart from the
\textit{ condition of continuity}. $f_1$ will be discontinuous when
$$
1=\alpha(3x_1+\rho)^{-1/3},~~~3x_1=\alpha^3-\rho.
$$
In order that this discontinuity coincides with the origin,
it must be
\begin{equation}\label{A13}
\rho=\alpha^3.
\end{equation}
Therefore the condition of continuity relates in this way the
two integration constants $\rho$ and $\alpha$.\par
The complete solution of our problem reads now:
$$
f_1=\frac{1}{R^4}\frac{1}{1-\alpha/R},~~
f_2=f_3=R^2,~~f_4=1-\alpha/R,
$$
where the auxiliary quantity
$$
R=(3x_1+\rho)^{1/3}=(r^3+\alpha^3)^{1/3}
$$
has been introduced.\par
    When one introduces these values of the functions
$f$ in the expression (\ref{A9}) of the line element and goes
back to the usual polar co-ordinates one gets \textit{ the line element
that forms the exact solution of Einstein's problem}:
\begin{equation}\label{A14}
\dd s^2=\left(1-{\alpha}/{R}\right)\dd t^2
-\frac{\dd R^2}{1-{\alpha}/{R}}
-R^2(\dd \vartheta^2+\sin^2\vartheta \dd \phi^2),~
R=(r^3+\alpha^3)^{1/3}.
\end{equation}
The latter contains only the constant $\alpha$, that depends on
the value of the mass at the origin.\par
\S 5. \textit{ The uniqueness of the solution} resulted
spontaneously through the present calculation. From what follows
we can see that it would have been difficult to ascertain the
uniqueness from an approximation procedure in the manner of
Mr. Einstein. Without the continuity condition it
would have resulted:
$$
f_1=\frac{(3x_1+\rho)^{-4/3}}{1-\alpha(3x_1+\rho)^{-1/3}}
=\frac{(r^3+\rho)^{-4/3}}{1-\alpha(r^3+\rho)^{-1/3}}.
$$
When $\alpha$ and $\rho$ are small, the series expansion up to
quantities of second order gives:
$$
f_1=\frac{1}{r^4}
\bigg[1+\frac{\alpha}{r}-\frac{4}{3}\frac{\rho}{r^3}\bigg].
$$
This expression, together with the corresponding expansions of $f_2$,
$f_3$, $f_4$, satisfies up to the same accuracy all the
conditions of the problem. Within this approximation the condition
of continuity does not introduce anything new, since discontinuities
occur spontaneously only in the origin. Then the two constants
$\alpha$ and $\rho$ appear to remain arbitrary, hence the problem
would be physically undetermined. The exact solution teaches that
in reality, by extending the approximations, the discontinuity does
not occur at the origin, but at
$r=(\alpha^3-\rho)^{1/3}$, and that one must set just
$\rho=\alpha^3$ for the discontinuity to go in the
origin. With the approximation in powers of $\alpha$ and $\rho$
one should survey very closely the law of the coefficients in
order to recognise the necessity of this link between $\alpha$ and
$\rho$.\par
\S 6. Finally, one has still to derive the \textit{ motion of a
point in the gravitational field}, the geodesic line
corresponding to the line element (\ref{A14}). From the three facts, that
the line element is homogeneous in the differentials and that its
coefficients do not depend on $t$ and on $\phi$, with the
variation we get immediately three intermediate integrals. If
one also restricts himself to the motion in the equatorial plane
($\vartheta=90^o$, $\dd \vartheta=0$) these intermediate integrals
read:
\begin{equation}\label{A15}
(1-\alpha/R)\bigg(\frac{\dd t}{\dd s}\bigg)^2 -\frac{1}{1-\alpha/R}
\bigg(\frac{\dd R}{\dd s}\bigg)^2-R^2\bigg(\frac{\dd \phi}{\dd s}\bigg)^2
={\rm const.}=h,
\end{equation}
\begin{equation}\label{A16}
R^2\frac{\dd \phi}{\dd s}={\rm const.}=c,
\end{equation}
\begin{equation}\label{A17}
(1-\alpha/R)\frac{\dd t}{\dd s}={\rm const.}=1 \ \
{\rm(determination~of~the~time~unit).}
\end{equation}
\noindent From here it follows
$$
\bigg(\frac{\dd R}{\dd \phi}\bigg)^2+R^2(1-\alpha/R)
=\frac{R^4}{c^2}[1-h(1-\alpha/R)]
$$
or with $1/R=x$
\begin{equation}\label{A18}
\bigg(\frac{\dd x}{\dd \phi}\bigg)^2=\frac{1-h}{c^2}+
\frac{h\alpha}{c^2}x-x^2+\alpha x^3.
\end{equation}
If one introduces the notations: ${c^2/h}=B$, ${(1-h)/h}=2A$,
this is identical to Mr. Einstein's equation (\ref{A11}),
\textit{ loc. cit.} and gives the observed anomaly of the perihelion of
Mercury.\par
Actually Mr. Einstein's approximation for the orbit goes into
the exact solution when one substitutes for $r$ the quantity
$$
R=(r^3+\alpha^3)^{1/3}
=r\bigg(1+\frac{\alpha^3}{r^3}\bigg)^{1/3}.
$$
Since $\alpha/r$ is nearly equal to twice the square of the
velocity of the planet (with the velocity of light as unit),
for Mercury the parenthesis differs from $1$ only for quantities
of the order $10^{-12}$. Therefore $r$ is virtually identical to
$R$ and Mr. Einstein's approximation is adequate to the
strongest requirements of the practice.\par
Finally, the exact form of the third Kepler's law for circular
orbits will be derived. Owing to (\ref{A16}) and (\ref{A17}), when one
sets $x=1/R$, for the angular velocity $n={\dd \phi/\dd t}$ it holds
$$
n=cx^2(1-\alpha x).
$$
For circular orbits both ${\dd x/\dd \phi}$ and ${\dd ^2x/\dd \phi^2}$ must
vanish. Due to (\ref{A18}) this gives:
$$
0=\frac{1-h}{c^2}+\frac{h\alpha}{c^2}x-x^2+\alpha x^3,~~
0=\frac{h\alpha}{c^2}-2x+3\alpha x^2.
$$
The elimination of $h$ from these two equations yields
$$
\alpha=2c^2x(1-\alpha x)^2.
$$
Hence it follows
$$
n^2=\frac{\alpha}{2}x^3=\frac{\alpha}{2R^3}
=\frac{\alpha}{2(r^3+\alpha^3)}.
$$
The deviation of this formula from the third Kepler's law is
totally negligible down to the surface of the Sun. For an
ideal mass point, however, it follows that the angular velocity does
not, as with Newton's law, grow without limit when the radius of
the orbit gets smaller and smaller, but it approaches a determined
limit
$$
n_0=\frac{1}{\alpha\sqrt{2}}.
$$
(For a point with the solar mass the limit frequency will be
around $10^4$ per second). This circumstance could be of interest,
if analogous laws would rule the molecular forces.

\section{From ``Grundlagen der Physik'': translation
of Hilbert's derivation of the field of a ``Massenpunkt''}\label{B}
\setcounter{equation}{41}
\noindent . . . . . . . . . . . . . . . . . . . . . . . . . . . . . . . .
 . . . . . . . . . . . . . . \par The integration
of the partial differential equations (36)
is possible also in another case, that for the first time has been dealt
with by Einstein\footnote[1]{Perihelbewegung des Merkur, Sitzungsber. d. Akad.
zu Berlin. 1915, 831.} and by Schwarz\-schild\footnote[2]{\"Uber das Gravitationsfeld
eines Massenpunktes, Sitzunsber. d. Akad. zu
Berlin. 1916, 189.}. In the following I provide for this case a
way of solution that does not make any hypothesis on the
gravitational potentials $g_{\mu\nu}$ at infinity, and that
moreover offers advantages also for my further investigations.
The hypotheses on the $g_{\mu\nu}$ are the following:
\begin{enumerate}
\item The interval is referred to a Gaussian coordinate system -
however $g_{44}$ will still be left arbitrary; i.e. it is\par
\smallskip
\centerline{$g_{14}=0, \ \ \ g_{24}=0, \ \ \ g_{34}=0.$}
\smallskip
\item The $g_{\mu\nu}$ are independent of the time coordinate $x_{4}$.
\item The gravitation $g_{\mu\nu}$ has central symmetry with
respect to the origin of the coordinates.
\end{enumerate}

According to Schwarzschild,
if one poses
\begin{eqnarray}\nonumber
w_{1}=r\cos\vartheta\\\nonumber
w_{2}=r\sin\vartheta \cos \varphi\\\nonumber
w_{3}=r\sin\vartheta \sin \varphi\\\nonumber
w_{4}=l
\end{eqnarray}
the most general interval corresponding to these hypotheses is represented
in spatial polar coordinates by the expression
\begin{equation} \label{B42}
F(r)\dd r^{2}+G(r)(\dd \vartheta^{2}+\sin^{2}\vartheta \dd \varphi^{2})+H(r)\dd l^{2},
\end{equation}
where $F(r)$, $G(r)$, $H(r)$ are still arbitrary functions of $r$.
If we pose
\[
r^{\ast}=\sqrt{G(r)},
\]
we are equally authorised to interpret $r^{\ast}$,
$\vartheta$, $\varphi$ as spatial polar coordinates. If we
substitute in (\ref{B42}) $r^{\ast}$ for $r$ and then drop the
symbol $\ast $, it results the expression
\begin{equation}
M(r)\dd r^{2}+r^{2}\dd \vartheta^{2}
+r^{2}\sin^{2}\vartheta \dd \varphi^{2}+W(r)\dd l^{2},
\label{B43}
\end{equation}
where $M(r)$, $W(r)$ mean the two essentially arbitrary functions of $r$.
The question is how the latter shall be determined in the most
general way, so that the differential equations (36) happen to be
satisfied.

To this end the known expressions $K_{\mu\nu}$, $K$, given in my
first communication, shall be calculated. The first step of this
task consists in writing the differential equations of the
geodesic line through variation of the integral
\[
\int {\left( M\left(\frac{\dd r}{\dd p}\right)^{2}+r^{2}\left(
\frac{\dd \vartheta}{\dd p}\right)^{2}+r^{2}\sin^{2}\vartheta\left(
\frac{\dd \varphi}{\dd p}\right)^{2}+W(\left(\frac{\dd l}{dp}\right)
^{2}\right) \dd p}.
\]
We get as Lagrange equations:
\begin{eqnarray}\nonumber
\frac{\dd^{2}r}{\dd p^{2}}+\frac{M^{\prime}}{2M}\left(\frac{\dd r}{\dd p}
\right)^{2}
-\frac{r}{M}\left[\left(\frac{\dd \vartheta}{\dd p}\right)^{2}
+\sin^{2}\vartheta\left(\frac{\dd \varphi}{\dd p}\right)^{2}\right]
-\frac{W^{\prime}}{2M}\left(\frac{\dd l}{\dd p}\right)^{2}=0,\\\nonumber
\frac{\dd^{2}\vartheta}{\dd p^{2}}+\frac{2}{r}\frac{\dd r}{\dd p}
\frac{\dd \vartheta}{\dd p}-\sin
\vartheta \cos\vartheta\left(\frac{\dd \varphi}{\dd p}\right)^{2}=0,\\\nonumber
\frac{\dd^{2}\varphi}{\dd p^{2}}+\frac{2}{r}\frac{\dd r}{\dd p}
\frac{\dd \varphi}{\dd p}+2\cot
\vartheta\frac{\dd \vartheta}{\dd p}\frac{\dd \varphi}{\dd p}=0,\\\nonumber
\frac{\dd^{2}l}{\dd p^{2}}+\frac{W^{\prime}}{W}\frac{\dd r}{\dd p}\frac{\dd l}{\dd p}=0;
\end{eqnarray}
here and in the following calculation the symbol $^{\prime}$ means
differentiation with respect to $r$. By comparison with the
general differential equations of the geodesic line:

\[
\frac{\dd^{2}w_{s}}{\dd p^{2}}+\sum_{\mu\nu}\left\{_{~s}^{\mu~\nu}\right\}
\frac{\dd w_{\mu}}{\dd p}\frac{\dd w_{\nu}}{\dd p}=0
\]
we infer for the bracket symbols $\left\{ _{~s}^{\mu~\nu}\right\}$
the following values (the vanishing ones are omitted):
\[
\left\{ _{~1}^{1~1}\right\} =\frac{1}{2}\frac{M^{\prime}}{M},\ \ \ \left\{
_{~1}^{2~2}\right\} =-\frac{r}{M},\ \ \ \left\{ _{~1}^{3~3}\right\}
=-\frac{r}{M}\sin^{2}\vartheta ,
\]
\[
\left\{ _{~1}^{4~4}\right\} =-\frac{1}{2}\frac{W^{\prime}}{M},\ \ \ \left\{
_{~2}^{1~2}\right\} =\frac{1}{r},\ \ \ \left\{ _{~2}^{3~3}\right\} =-\sin\vartheta
\cos\vartheta ,
\]
\[
\left\{ _{~3}^{1~3}\right\} =\frac{1}{r},\ \ \ \left\{ _{~3}^{2~3}\right\}
=\cot\vartheta, \ \ \ \left\{ _{~4}^{1~4}\right\}
=\frac{1}{2}\frac{W^{\prime}}{W}.
\]
With them we form:
\[
K_{11}=\frac{\partial}{\partial r}\left(\left\{ _{~1}^{1~1}\right\} +\left\{
_{~2}^{1~2}\right\} +\left\{ _{~3}^{1~3}\right\} +\left\{ _{~4}^{1~4}\right\}
\right) -\frac{\partial}{\partial r}\left\{ _{~1}^{1~1}\right\}
\]
\[
+\left\{ _{~1}^{1~1}\right\}\left\{ _{~1}^{1~1}\right\} +\left\{
_{~2}^{1~2}\right\}\left\{ _{~2}^{2~1}\right\} +\left\{ _{~3}^{1~3}\right\}
\left\{ _{~3}^{3~1}\right\} +\left\{ _{~4}^{1~4}\right\}\left\{
_{~4}^{4~1}\right\}
\]
\[
-\left\{ _{~1}^{1~1}\right\}\left(\left\{ _{~1}^{1~1}\right\}
+\left\{ _{~2}^{1~2}\right\} +\left\{ _{~3}^{1~3}\right\} +\left\{
_{~4}^{1~4}\right\} \right)
\]
\[
=\frac{1}{2}\frac{W^{\prime \prime}}{W}+\frac{1}{4}\frac{W^{\prime 2}}{W^{2}}
-\frac{M^{\prime}}{rM}-\frac{1}{4}\frac{M^{\prime}W^{\prime}}{MW}
\]
\smallskip
\[
K_{22}=\frac{\partial}{\partial\vartheta}\left\{ _{~3}^{2~3}\right\}
-\frac{\partial}{\partial r}\left\{ _{~1}^{2~2}\right\}
\]
\[
+\left\{ _{~2}^{2~1}\right\}\left\{ _{~1}^{2~2}\right\} +\left\{
_{~1}^{2~2}\right\}\left\{ _{~2}^{1~2}\right\} +\left\{ _{~3}^{2~3}\right\}
\left\{ _{~3}^{3~2}\right\}
\]
\[
-\left\{ _{~1}^{2~2}\right\}\left(\left\{
_{~1}^{1~1}\right\}+\left\{ _{~2}^{1~2}\right\} +\left\{
_{~3}^{1~3}\right\} +\left\{ _{~4}^{1~4}\right\} \right)
\]
\[
=-1-\frac{1}{2}\frac{rM^{\prime}}{M^{2}}+\frac{1}{M}+\frac{1}{2}\frac{%
rW^{\prime}}{MW}
\]
\smallskip
\[
K_{33}=-\frac{\partial}{\partial r}\left\{ _{~1}^{3~3}\right\} -\frac{%
\partial}{\partial\vartheta}\left\{ _{~2}^{3~3}\right\}
\]
\[
+\left\{ _{~3}^{3~1}\right\}\left\{ _{~1}^{3~3}\right\} +\left\{
_{~3}^{3~2}\right\}\left\{ _{~2}^{3~3}\right\} +\left\{ _{~1}^{3~3}\right\}
\left\{ _{~3}^{1~3}\right\} +\left\{ _{~2}^{3~3}\right\}\left\{
_{~3}^{2~3}\right\}
\]
\[
-\left\{ _{~1}^{3~3}\right\}\left(\left\{ _{~1}^{1~1}\right\}
+\left\{ _{~2}^{1~2}\right\} +\left\{ _{~3}^{1~3}\right\} +\left\{
_{~4}^{1~4}\right\} \right)-\left\{ _{~2}^{3~3}\right\}\left\{
_{~3}^{2~3}\right\}
\]
\[
=\sin^{2}\vartheta\left( -1-\frac{1}{2}\frac{rM^{\prime}}{M^{2}}+\frac{1}{M}+\frac{%
1}{2}\frac{rW^{\prime}}{MW}\right)
\]
\smallskip
\[
K_{44}=-\frac{\partial}{\partial r}\left\{ _{~1}^{4~4}\right\} +\left\{
_{~4}^{4~1}\right\}\left\{ _{~1}^{4~4}\right\} +\left\{ _{~1}^{4~4}\right\}
\left\{ _{~4}^{4~1}\right\}
\]
\[
-\left\{ _{~1}^{4~4}\right\}\left(\left\{ _{~1}^{1~1}\right\}
+\left\{ _{~2}^{1~2}\right\} +\left\{ _{~3}^{1~3}\right\} +\left\{
_{~4}^{1~4}\right\} \right)
\]
\[
=\frac{1}{2}\frac{W^{\prime \prime}}{M}-\frac{1}{4}\frac{M^{\prime
}W^{\prime}}{M^{2}}-\frac{1}{4}\frac{W^{\prime 2}}{MW}+\frac{W^{\prime}}{rM%
}
\]
\smallskip
\[
K=\sum_{s}g^{ss}K_{ss}=\frac{W^{\prime \prime}}{MW}-\frac{1}{2}%
\frac{W^{\prime 2}}{MW^{2}}-2\frac{M^{\prime}}{rM^{2}}-\frac{1}{2}\frac{%
M^{\prime}W^{\prime}}{M^{2}W}-\frac{2}{r^{2}}+\frac{2}{r^{2}M}+2\frac{%
W^{\prime}}{rMW}.
\]
Since
\[
\sqrt{g}=\sqrt{MW}r^{2}\sin\vartheta
\]
it is found
\[
K\sqrt{g}=\left\{\left(\frac{r^{2}W^{\prime}}{\sqrt{MW}}\right)^{\prime
}-2\frac{rM^{\prime}\sqrt{W}}{M^{\frac{3}{2}}}-2\sqrt{MW}+2\sqrt{\frac{W}{M}%
}\right\} \sin\vartheta \
\]
and, if we set
\[
M=\frac{r}{r-m},\ \ \ \ W=w^{2}\frac{r-m}{r},\
\]
where henceforth $m$ and $w$ become the unknown functions of $r$,
we eventually obtain
\[
K\sqrt{g}=\left\{\left(\frac{r^{2}W^{\prime}}{\sqrt{MW}}\right)^{\prime}
-2wm^{\prime}\right\} \sin\vartheta .\
\]
Therefore the variation of the quadruple integral
\[
\int\int\int\int K\sqrt{g}\dd r\dd \vartheta \dd \varphi \dd l\
\]
is equivalent to the variation of the single integral
\[
\int wm^{\prime}\dd r\
\]
and leads to the Lagrange equations
\begin{eqnarray}\label{B44}
m^{\prime}=0,\\\nonumber
w^{\prime}=0.
\end{eqnarray}
One easily satisfies oneself that these equations effectively entail the
vanishing of all the $K_{\mu\nu}$; they represent therefore
essentially the most general solution of the equations (36) under
the hypotheses (1), (2), (3) previously made.
 If we take as integrals of (\ref{B44}) $m=\alpha$,
where $\alpha$ is a constant, and $w=1$ (a choice that evidently does
not entail any essential restriction)
from (\ref{B43}) with $l=it$ it results the looked for interval
in the form first found by Schwarzschild

\begin{equation}
G(\dd r,\dd \vartheta ,\dd \varphi ,\dd l)
=\frac{r}{r-\alpha}\dd r^{2}+r^{2}\dd \vartheta
^{2}+r^{2}\sin^{2}\vartheta \dd \varphi^{2}-\frac{r-\alpha}{r}\dd t^{2}.
\label{B45}
\end{equation}
The singularity of this interval for $r=0$ vanishes only when it is
assumed $\alpha=0$, i.e.: {\gr under the hypotheses (1), (2), (3)
the interval of the pseudo-Euclidean geometry is the only regular interval
that corresponds to a world without electricity.}

For $\alpha\neq 0$,  $r=0$ and, with positive values of $\alpha$, also
$r=\alpha$ happen to be such points that in them the interval is not regular.
I call an interval or a gravitational field $g_{\mu\nu}$ \textit{ regular}
in a point if, through an invertible one-to-one transformation, it is possible
to introduce a coordinate system such that for it the
corresponding functions $g_{\mu\nu}^{\prime}$ are regular in that
point, i.e. in it and in its neighbourhood they are
continuous and differentiable at will, and have a determinant
$g^{\prime}$ different from zero.

Although in my opinion only regular solutions of the fundamental
equations of physics immediately represent the reality,
nevertheless just the solutions with non regular points are an
important mathematical tool for approximating characteristic
regular solutions - and in this sense, according to the procedure
of Einstein and Schwarzschild, the interval (\ref{B45}), not regular
for $r=0$ and for $r=\alpha$, must be considered as expression
of the gravitation of a mass distributed with central symmetry
in the surroundings of the origin\footnote[1]{Transforming to the origin
the position $r=\alpha$, like Schwarzschild did, is in my opinion not
advisable; moreover Schwarz\-schild's transformation is not the
simplest one, that reaches this scope.}. . . . . . . . . . .
. . . . . . . . . . . . . . . . . . . . . .
 . . . . . . .


\newpage


\begin{thebibliography}{}

\bibitem{Schwarzschild1916} Schwarzschild, K. (1916). { Sitzungsber. Preuss.
Akad. Wiss., Phys. Math. Kl.}, 189 (submitted 13 Jan. 1916).

\bibitem{Abrams1989} Abrams, L.S. (1989). { Can. J. Phys.} {\bf 67}, 919.
\par\noindent  http://arxiv.org/abs/gr-qc/0102055.

\bibitem{Hilbert1917} Hilbert, D. (1917). { Nachr. Ges. Wiss. G\"ottingen,
Math. Phys. Kl.}, 53 (submitted 23 Dec. 1916).

\bibitem{Brillouin1923} Brillouin, M., (1923). { J. Phys. Rad.}
{\bf 23}, 43.\par\noindent
English translation at: http://arxiv.org/abs/physics/0002009.

\bibitem{Schwarzschild2003} English translation of
\cite{Schwarzschild1916}: (2003). { Gen. Relativ. Gravit.} {\bf 35}, 951.
\par\noindent  http://arXiv.org/abs/physics/9905030.

\bibitem{Note2003} Antoci, S., and Liebscher, D.-E. (2003). { Gen.
Relativ. Gravit.} {\bf 35}, 945.

\bibitem{Einstein1915c} Einstein, A. (1915). { Sitzungsber. Preuss.
Akad. Wiss., Phys. Math. Kl.}, 844 (submitted 25 Nov. 1915).

\bibitem{Hilbert1915} Hilbert, D. (1915) { Nachr. Ges. Wiss. G\"ottingen,
Math. Phys. Kl.}, 395 (submitted 20 Nov. 1915).

\bibitem{Einstein1915a} Einstein, A. (1915). { Sitzungsber. Preuss.
Akad. Wiss., Phys. Math. Kl.}, 778, 799 (submitted 11 Nov. 1915).

\bibitem{Mie1912} Mie, G. (1912). { Annalen der Physik}
{\bf 37}, 511; {ibidem} {\bf 39}, 1.

\bibitem{Mie1913} Mie, G. (1913). { Annalen der Physik}
{\bf 40}, 1.

\bibitem{Einstein15b} Einstein, A. (1915). { Sitzungsber. Preuss.
Akad. Wiss., Phys. Math. Kl.}, 831 (submitted 18 Nov. 1915).

\bibitem{Lichnerowicz1955} Lichnerowicz, A., (1955). \textit{ Th\'eories
relativistes de la gravitation et de l'\'e\-lectro\-magn\'etisme}, Masson,
Paris.

\bibitem{Einstein1916} Einstein, A. (1916). { Annalen der Physik}
{\bf 49}, 769.

\bibitem{Eddington1924} Eddington, A.S., (1924). { Nature} {\bf 113}, 192.

\bibitem{Finkelstein1958} Finkelstein, D.,  (1958). { Phys. Rev.}
{\bf 110}, 965.

\bibitem{Lemaitre1933} Lemaitre, G., (1933). { Ann. Soc. Sci.
Bruxelles} {\bf 53}, 51.

\bibitem{Synge1950} Synge, J.L., (1950). { Proc. R. Irish Acad.}
{\bf 53A}, 83.

\bibitem{Fronsdal1959} Fronsdal, C., (1959). { Phys. Rev.} {\bf 116}, 778.

\bibitem{Kruskal1960} Kruskal, M.D., (1960). { Phys. Rev.} {\bf 119}, 1743.

\bibitem{Szekeres1960} Szekeres, G., (1960). { Publ. Math. Debrecen} {\bf 7}, 285.

\bibitem{Rindler2001} Rindler, W., (2001). \textit{ Relativity, special, general
and cosmological}, Oxford University Press, Oxford, pp. 265-267.

\bibitem{Geroch1968a} Geroch, R., (1968). {J. Math. Phys.}{\bf 9}, 450.

\bibitem{Geroch1968b} Geroch, R., (1968). {Annals of Physics} {\bf 48}, 526.

\bibitem{Schmidt1971} Schmidt, B.G., (1971). { Gen. Relativ. Gravit.} {\bf 1}, 269.

\bibitem{GKP1972} Geroch, R., Kronheimer, E.H., and Penrose, R.
(1972). { Proc. R. Soc. Lond. A} {\bf 327}, 545.

\bibitem{ES1977} Ellis, G.F.R., and Schmidt, B.G. (1977).
{ Gen. Relativ. Gravit.} {\bf 8}, 915.

\bibitem{Thorpe1977} Thorpe, J.A., (1977). { J. Math. Phys.}
{\bf 18}, 960.

\bibitem{GLW1982} Geroch, R., Liang Can-bin, and Wald, R.M., (1982).
{ J. Math. Phys.} {\bf 23}, 432.

\bibitem{SS1994} Scott, Susan M., and Szekeres, P., (1994).
{ J. Geom. Phys.} {\bf 13}, 223.
\par\noindent http://arxiv.org/abs/gr-qc/9405063.

\bibitem{Synge1966} Synge, J. L., (1966). \textit{ What is Einstein's Theory of
Gravitation?}, in: Hoffman, B. (ed.),
\textit{ Essays in Honor of Vaclav Hlavat\'y}, Indiana University Press,
Bloomington p. 7.

\bibitem{Israel1967a} Israel, W., (1967). { Phys. Rev.} {\bf 164},
1776.

\bibitem{BW1922} Bach, R. and Weyl, H., (1922). { Math. Zeitschrift}
{\bf 13}, 134.

\bibitem{Weyl1917} Weyl, H., (1917). { Annalen der Physik} {\bf 54},
 117.

\bibitem{Levi-Civita1919} Levi-Civita, T., (1919). { Rend. Acc. dei Lincei}
{\bf 28}, 3.

\bibitem{Zipoy1966} Zipoy, D.M., (1966). { J. Math. Phys.} {\bf 7},
 1137.

\bibitem{CJ1973} Cooperstock, F.I. and Junevicus, G.J., (1973).
{ Nuovo Cimento B} {\bf 16}, 387.

\bibitem{Virbhadra1996} Virbhadra, K.S., (1996). { e-print},
http://arXiv.org/abs/gr-qc/9606004.

\bibitem{Israel1967b} Israel, W., (1967). { Nature} {\bf 216},
148.

\bibitem{Synge1937} Synge, J. L., (1937). { Proc. Math. Soc. Edinburgh}
{\bf 5}, 93.

\bibitem{Whittaker1935} Whittaker, E. T., (1935). { Proc. R. Soc. London A}
{\bf 149}, 384.

\bibitem{PU1955} Papapetrou, A. and Urich, W., (1955). { Z. Naturforschg. A}
{\bf 10}, 109.

\bibitem{Synge1934} Synge, J. L., (1934). { Ann. Math.} {\bf 35}, 705.

\bibitem{Levi-Civita1918} Levi-Civita, T., (1918). { Rend. Acc. dei Lincei}
{\bf 27}, 3.

\bibitem{ALM2003} Antoci, S., Liebscher, D.-E., and Mihich, L., (2003).
{ Astron. Nachr.} {\bf 324}, 485.
\par\noindent http://arxiv.org/abs/gr-qc/0107007.

\bibitem{ALM2001} Antoci, S., Liebscher, D.-E., and Mihich, L., (2001).
{ Class. Quantum Grav.} {\bf 18}, 3463.
\par\noindent http://arxiv.org/abs/gr-qc/0104035.

\bibitem{Weyl1919} Weyl, H., (1919). { Annalen der Physik} {\bf 59},
185.

\bibitem{Combridge1923} Combridge, J.T., (1923). { Phil. Mag.} {\bf 45}, 726.

\bibitem{Janne1923} Janne, H., (1923). { Bull. Acad. R. Belg.} {\bf 9}, 484.

\bibitem{de Sitter1916} de Sitter, W., (1916). { Month. Not. R. Astr. Soc.}
{\bf 76}, 699.

\bibitem{Abrams1979} Abrams, L.S., (1979). { Phys. Rev. D }{\bf 20}, 2474.
\par\noindent  http://arxiv.org/abs/gr-qc/0201044.

\end{thebibliography}
\end{document}